\documentclass[useAMS,usenatbib,usegraphicx]{mn2e}

\newcommand{\deltac}{\delta_{\rm c}}
\newcommand{\deltasc}{\delta_{\rm sc}}
\newcommand{\deltaeff}{\delta_{\rm eff}}
\newcommand{\deltal}{\delta_{\rm l}}

\newcommand{\meanrho}{\overline{\rho}}
\newcommand{\runit}{ {\rm h}^{-1}~{\rm Mpc}}
\newcommand{\munit}{ {\rm h}^{-1}~{\rm M}_\odot}
\newcommand{\ncavg}{n_c^{\rm avg}}
\newcommand{\mstar}{ {\rm m}_{\star} }

\title[CMF of dark matter haloes]{A prescription for the conditional mass
function of dark matter haloes}

\author[J.~A. Rubi\~no-Mart\'{\i}n et al.]{
Jos\'e~Alberto Rubi\~no-Mart\'{\i}n$^{1}$\thanks{E-mail:
jose.alberto.rubino@iac.es}, 
Juan Betancort-Rijo$^{1,2}$\thanks{E-mail: jbetanco@iac.es} and 
Santiago G. Patiri$^{1}$\thanks{Present address: Department of Astronomy, 
Case Western Reserve University
10900 Euclid Ave., Cleveland, Ohio, 44106, USA}\\
$^{1}$Instituto de Astrof\'{\i}sica de Canarias, C/V\'{\i}a L\'actea s/n,
    E-38200 Tenerife, Spain\\
$^{2}$Facultad de F\'{\i}sicas, Universidad de la Laguna, C/Astrof\'{\i}sico
Francisco  S\'anchez s/n, , E-38200 Tenerife, Spain
}

\begin{document}

\date{}

\pagerange{\pageref{firstpage}--\pageref{lastpage}} \pubyear{2007}

\maketitle

\label{firstpage}

\begin{abstract}
The unconditional mass function (UMF) of dark matter haloes has been determined
accurately in the literature, showing excellent agreement with high resolution
numerical simulations.  However, this is not the case for the conditional mass
function (CMF).
%%%
Here, we propose a simple analytical procedure to derive the CMF by rescaling
the UMF to the constrained environment using the appropriate mean and variance
of the density field at the constrained point.
This method introduces two major modifications with respect to the standard
re-scaling procedure. First of all, rather than using in the scaling procedure
the properties of the environment averaged over all the conditioning region, we
implement the re-scaling locally. We show that for high masses this modification
may lead to substantially different results.
Secondly, we modify the (local) standard re-scaling procedure in such a manner
as to force normalisation, in the sense that when one integrates the CMF over
all possible values of the constraint multiplied by their corresponding
probability distribution, the UMF is recovered.
In practise, we do this by replacing in the standard procedure the value
$\deltac$ (the linear density contrast for collapse) by certain adjustable
effective parameter $\deltaeff$.
In order to test the method, we compare our prescription with the results
obtained from numerical simulations in voids \citep{Gottlober}.  We find that
when our modified re-scaling is applied locally to any existing numerical fit of
the UMF, and the appropriate value for $\deltaeff$ is chosen, the resulting CMF
is, in all cases, in very good agreement with the numerical results.
Based on these results, we finally present a very accurate
analytical fit to the (accumulated) conditional mass function obtained
with our procedure, as a function of the parameters that describe the
conditioning region (size and mean linear density contrast), the
redshift and the relevant cosmological parameters ($\sigma_8$ and
$\Gamma$). This analytical fit may be useful for any theoretical
treatment of the large scale structure, and has been already used
successfully in regard with the statistic of voids.
\end{abstract}

\begin{keywords}
methods: analytical -- methods: statistical -- cosmology: theory --
dark matter -- large-scale structure of the Universe
\end{keywords}

%----------------------------------
\section{Introduction}

In the last few years, there has been considerable effort in obtaining
accurate theoretical predictions for the mass function of collapsed
dark matter haloes.
%----------
% Decir ejemplos de aplicacion
%----------
By far, the most widely used prediction for the unconditional mass function
(UMF) is the \citet[][hereafter PS]{PS} formalism. Extensions of this
prescription provide a way to compute not only good approximations to the UMF
\citep{PS,Bond91}, but also the merging history \citep{Bond91,LaceyCole93} and
the spatial clustering \citep{MoWhite,ShethLemson99} of dark matter haloes.

The two basic assumptions of the PS approach are the physics of the
spherical collapse, and the fact that the initial fluctuations were
drawn from a gaussian distribution.
\citet{Bond91} showed how to combine these two hypothesis in order to
obtain the UMF of dark matter haloes from the barrier crossing
statistics of many independent, uncorrelated random walks (the
so-called ``excursion set formalism'').  The barrier shape is given by
the fact that, in the spherical collapse model, a certain region
collapses at time $t$ (or redshift $z$) if the initial overdensity
within it exceeds a critical value ($\deltasc$) which is independent
of mass.
However, numerical simulations show that the PS UMF as derived from
this formalism, while qualitatively correct, overestimates the
abundance of ``typical'' haloes and underestimates that of massive
ones when compared with the results of N-body simulations
\citep[e.g.][]{EFWD88}.

In order to understand if this discrepancy could be due to the
assumption of spherical collapse made within the PS formalism, several
works in the last few years extended the excursion set formalism by
incorporating a treatment of the ellipsoidal collapse
\cite[][]{BondMyers96,S01,S02}. In practise, this is done by using a
``moving barrier'', i.e. a barrier whose height depends on mass. This
study provides an analytic expression which describes to a good
approximation the distribution of first-crossings of this ``moving''
barrier, once two free parameters are fitted by comparison with N-body
simulations \citep[][ST]{ST}.
The resulting mass function significantly improves the results of the
PS formalism, although a small discrepancy still remains at high
masses \citep{J01}. The authors of this last paper also propose an
analytic fitting formula which better reproduces the numerical
results, but which can not be extrapolated in mass beyond the range of
the fit.

Recently, \citet[][WA]{WA} use a large sample of simulations to
provide a extremely good fitting formula for the UMF at redshift zero,
while \citet{Reed07} provide a good fit to the UMF at redshifts 10-30.
On the other hand, \cite{BM,BM2} proposed an analytical procedure for deriving
the UMF at any redshift, which also includes the physics of the ellipsoidal
collapse encoded in the ST mass function.  In this procedure, the
``all-mass-at-center'' problem is treated in an appropriate manner. As a
consequence of this, the high mass behaviour and the redshift dependence of the
UMF comes out right without introducing any fudge parameter to be fitted with
simulations.  Their expression reproduces well the \citet{WA} fit to the
simulations.

Summarising, there is a variety of analytical expressions which accurately
reproduce the UMF. However, this is not the case for the conditional mass
function (CMF).
% History
There are different approaches in the literature to obtain an analytic
expression for the CMF of dark matter haloes, $n_c(m)$.  The most widely used
framework is again the {\it excursion set formalism}, often called {\it extended
Press-Schechter} \citep[][]{Bond91,LaceyCole93}, and denoted EPS \citep[for
a recent review, see][]{2007IJMPD..16..763Z}. 
However, this approach does not compares well to N-body simulations \citep[see
e.g.][]{Tormen98}, being one of the reasons that it does not include the physics
of the ellipsoidal collapse. \citet{S02} propose an alternative expression of
the CMF to solve this issue within the context of the ellipsoidal collapse
moving barrier, which reproduces much better the results from simulations of the
hierarchical assembly of dark matter haloes.

% OUR PROPOSAL:
In this paper, we present our method to build the CMF of dark matter haloes.
Our motivation is to provide an analytical prescription to build the CMF as an
extension of the UMF, taking advantage of the high degree of accuracy at which
we know these functions.
Our proposal consists in a prescription to rescale the UMF, which can be applied
to any of the existing fitting formula, and which takes into account a certain
normalisation condition that we will describe below.
Our procedure differs from the standard one in two major respects. First, the
scaling of the UMF is implemented locally, using at each point within the
conditioning region the local environment, rather than using the average (in
practise, this means that our CMF will depend on an additional variable, $q$,
the distance from the center of the condition to the point at which we evaluate
the CMF, and over which we have to integrate in order to obtain the
average CMF).  Secondly, the scaling procedure is somewhat different from the
standard one, and instead of the usual value of $\deltac$, we will use a
modified one.

We assume throughout the paper a flat $\Lambda$CDM model with the following
cosmological parameters: $\Omega_{\rm m} = 0.3$, $h=0.7$
(i.e. $\Gamma=\Omega_{\rm m} h=0.21$),
$\sigma_8=0.9$ and no tilt of the primordial spectrum (i.e. a scalar spectral
index $n_S$ of $1$).
The CDM power spectrum ($P(k) \propto k^{n_{\rm S}} T(k)^2$) adopted for our
computations is obtained from the transfer function $T(k)$ given by
\cite{Bond84},
\begin{equation}
T(k) = \Big( 1 + [ak + (bk)^{3/2} + (ck)^2]^\nu \Big)^{-1/\nu}
\end{equation}
where $a = 6.4/\Gamma$~$\runit$, $b = 3.0/\Gamma$~$\runit$, $c =
1.7/\Gamma$~$\runit$, and $\nu=1.13$.
%

%%%%%%%%%%%%%%%%%%%%%%%%%%%%%%%%%%%%%%%%%%%%%%%%%%%
%%%%%%%%%%%%%%%%%%%%%%%%%%%%%%%%%%%%%%%%%%%%%%%%%%%
%%%%%%%%%%%%%%%%%%%%%%%%%%%%%%%%%%%%%%%%%%%%%%%%%%%
\section{The unconditional mass function}

The (unconditional) mass function (UMF) of dark matter haloes at a given
redshift $z$, $n(m,z)$, is defined such that $n(m,z) dm$ is the comoving number
density of haloes in the mass range $(m, m+ dm)$ at that redshift. The bound
objects (haloes) are usually defined in numerical simulations using two
different algorithms: the friends-of-friends finder \citep{Davis85}, and the
spherical overdensity finder \citep{LaceyCole94}.  Note that different
algorithms may lead to differences of the order of 10-20\% in the mass functions
\citep[see the discussion in][]{J01}.

The UMF is related with $F(m,z)$, the mass fraction in collapsed objects with
masses greater or equal than $m$, as
\begin{equation}
n(m,z) = - \frac{\meanrho}{m} \frac{d F(m,z)}{d m},
\end{equation}
where $\meanrho$ is the mean matter density of the Universe. 
It is also common to write the mass dependence in terms of $\sigma (m,z)$, where
$\sigma^2 (m,z)$ denote the mass variance of the linear density field
extrapolated to the redshift $z$ at which the haloes are defined, and on mass
scale $m \propto \meanrho r^3$ (the precise form of the $m-r$ relation depends
on the window function used).  This variance can be obtained as
\begin{equation}
\sigma^2 (m,z) = \frac{b(z)}{2\pi^2} \int_{0}^{+\infty} |\delta_k|^2 
W^2(k r) k^2 dk 
\end{equation}
where $|\delta_k|^2$ is the linear power spectrum of density fluctuations, and
$b(z)$ is the growth factor of linear perturbations normalised to unity at $z=0$
\citep{Lahav91,Carroll92}.
The window function is usually taken to be a real-space top-hat filter, i.e.
$m=4\pi \meanrho r^3 / 3$, and
\begin{equation}
W( x )= \frac{3}{x^3} (\sin x - x \cos x)
\end{equation}

Using $\ln \sigma^{-1}$ as the mass variable, the UMF is also defined
in some works as
\begin{equation}
f(\sigma,z) = \frac{m}{\meanrho} \frac{d N(m,z)}{d \ln \sigma^{-1}}
\label{eq:f}
\end{equation}
where now $f$ represents the mass fraction contained in collapsed objects per
unit $\ln \sigma^{-1}$, and $N(m,z)$ is the number density of collapsed objects
with masses above $m$, i.e.
\begin{equation}
N(m,z) = \int_{m}^{+\infty} n(m',z) dm' .
\end{equation}
The $f$ function is connected with the standard $F(m,z)$ as
\begin{equation}
  F(\sigma,z) = \int_{-\infty}^{ \ln \sigma} f(\sigma',z) d\ln \sigma' 
\label{eq:F_and_f}
\end{equation}

The advantage of using eq.~\ref{eq:f} is that the majority of the analytic (or
semi-analytic) models for the UMF predict a functional form for $f(\sigma,z)$
with no explicit dependence on redshift.
In those cases, all the dependence in mass and redshift can be absorbed
into a single variable, $\nu = (\deltac / \sigma(m,z) )^2$, where $\deltac$ is
the threshold parameter usually taken to be the extrapolated linear overdensity
$\deltal$ of a spherical perturbation at the time it collapses.  For the case of
an Einstein-de Sitter cosmology, we have $\delta_c = 1.686$
\cite[e.g][]{Peebles80}, with a weak dependence on the cosmological parameters
\citep[e.g][]{NFW}.

%
% UMF used here. 
In this paper, we shall consider four of the different prescriptions for the UMF
which were mentioned in the introduction, namely PS, ST, WA and BM.  However,
the method described here can be applied in principle to any other
prescription\footnote{As we will see in Sec.~\ref{sec:normalization}, in order
to use the proposed normalisation condition, it is required that the mass
function should have appropriate asymptotic behaviours at high and low
masses. For this reason, we can not use here the \citet{J01} UMF. }.
The expressions for these UMF are the following:
\begin{enumerate}
\item[(i)] The Press-Schechter mass-function \citep{PS,Bond91,LaceyCole93} is given by
\begin{equation}
  f_{\rm PS}(\sigma) = \sqrt{\frac{2}{\pi}} \frac{\delta_c}{\sigma}
  \exp ( - \frac{\delta_c^2}{2\sigma^2} )
\end{equation}
%
%ST mass function
%
\item[(ii)] The empirical best-fit mass function of ST is given by
\begin{equation}
  f_{\rm ST}(\sigma) = A \sqrt{\frac{2a}{\pi}}  
\Big[ 1+ \Big( \frac{\sigma^2}{2 \delta_c^2} \Big)^p \Big] \frac{\delta_c}{\sigma}
\exp ( - \frac{a\delta_c^2}{2\sigma^2})
\label{eq:f_st}
\end{equation}
where $A=0.3222$, $a=0.707$ and $p=0.3$. 
%
% Warren mass function
\item[(iii)] The \citet{WA} UMF uses a modified version 
of the functional form proposed by ST which is fitted to the simulations, giving
\begin{equation}
  f_{\rm WA}(\sigma) = A ( \sigma^{-a} + b ) \exp ( - \frac{c}{\sigma^2})
\label{eq:f_wa}
\end{equation}
with $A=0.7234$, $a=1.625$, $b=0.2538$ and $c=1.1982$. 
This expression provides a very good fit, over a mass range of $5$ orders of
magnitude, to the UMF obtained from simulations.
%
% BM mass function
%
\item[(iv)] The \citet{BM} mass function does not depends only on $\nu$, but
there is an additional dependence on the mass which enters through certain local
spectral index, which is characterised by a parameter represented by $c(m)$. In
this case, the UMF is given by
\begin{equation}
  f_{\rm BM}(\sigma, c(m)) = \frac{d F_{\rm BM}(m)}{d \ln \sigma } 
\end{equation}
where the mass fraction $F_{\rm BM}(m)$ is defined as
\begin{equation}
  F_{\rm BM}(m) \equiv \frac{ F_{\rm ST,a=1}(m) }{ V(m) }, 
\end{equation}
\begin{equation}
V(m) = 3 \int_{0}^{1} {\rm erfc} \Big[ \frac{\delta_c}{\sqrt{2}\sigma}
\Big(\frac{1 - \exp(-c(m) u^2)}{1 + \exp( -c(m) u^2)}\Big)^{1/2} \Big] u^2 du
\end{equation}
and $F_{\rm ST,a=1}(m)$ is the ST mass fraction which corresponds to
equation~\ref{eq:f_st} but using $a=1$.
The expression for $c(m)$ will be discussed in Section~\ref{sec:rescaling}
(equations \ref{eq:Dq}, \ref{eq:cm} and \ref{eq:dcm}).
\end{enumerate}

Given that the BM mass function is defined in terms of the mass fraction $F(m)$,
we find it useful to have analytical expressions for the mass fractions of the
different UMF. These formulae are presented in Appendix~\ref{app1}.

%%%%%%%%%%%%%%%%%%%
%%%%%%%%%%%%%%%%%%%

\section{The conditional mass function}

% History. Conditional MFs.
As discussed above, one can find several analytic expressions in the literature
for the conditional mass function (CMF) of dark matter
haloes, $n_c(m)$. The most widely used framework is the so
called {\it extended Press-Schechter (EPS)}, or excursion set formalism, either
in the context of spherical collapse \citep{Bond91,LaceyCole93} or in a more
elaborated form, in the case of a constant barrier.  According to the later
formalism, the CMF can be estimated by considering successive crossings of
barriers of different heights.
In the case of constant barriers this problem has an analytic solution, and it
is possible to provide analytic expressions for both the UMF and the CMF. In
particular, the CMF essentially has the same form as the UMF, but in the
rescaled variables
\begin{eqnarray}
\nonumber
\deltac   \rightarrow & \deltac - \delta_{0}\\
\sigma^2  \rightarrow & \sigma^2 - \sigma^2_0, 
\label{eq:standard}
\end{eqnarray}
where $\delta_{0}$ and $\sigma^2_0$ are the linear density and the
amplitude of the condition.
Hereafter, we will refer to this equation as the standard
re-scaling. Note that this is precisely the scaling which is obtained
for the conditional probability distribution of the average density
field within scale $\sigma^2$ at a randomly chosen point
within a region on scale $\sigma_0^2$ with inner linear density
contrast $\delta_0$ \citep[see appendix A of][]{Bower91}.
Within this EPS formalism, the CMF would be given
by\footnote{Note that according to this definition, the
conditional mass function $n_{\rm c}$ is defined as ``lagrangian'',
i.e. we explicitly use the mean density, $\meanrho$, when defining the
volume element, and not the actual local density in the conditional
region.  Throughout this paper, we will always refer to lagrangian
CMFs unless otherwise stated. }
\[
n_{\rm c, EPS}(m | \delta_{0}, \sigma_0) = \Big( \frac{2}{\pi} \Big)^{1/2}
\frac{\meanrho}{m} \Big| \frac{d \sigma}{d m} \Big| 
\frac{\sigma (\delta_c -  \delta_{0})}{ (\sigma^2 - \sigma^2_0 )^{3/2}} 
\times
\]
\begin{equation}
\label{eq:eps}
\qquad \qquad 
\exp \Bigg\{ - \frac{(\deltac-\delta_{0})^2}{2(\sigma^2 -\sigma_0^2)}  \Bigg\}
\end{equation}
However, as discussed in the last section, the EPS is not the most
accurate approximation to the halo abundance, being one of the reasons
that it does only contains the physics of the spherical collapse.
%

% ST02
%
The \citet[][hereafter ST02]{S02} approximation for the CMF further
extends this EPS formalism by including the physics of the ellipsoidal
collapse. This is done by using the moving barrier shape derived by
\citet{S01}, which is given by
\begin{equation}
B_{\rm ec}(\sigma^2,z) = \sqrt{a} \deltac(z) 
\Big[ 1 + \beta  \Big( \frac{a\deltac(z)^2}{\sigma^2} \Big)^{-\alpha} \Big],
\label{eq:barrier_ec}
\end{equation} 
with $a=0.707$, $\beta=0.485$ and $\alpha=0.615$. Note that with our notation,
the redshift dependence on the growth factor is contained inside $\sigma(m,z)$,
so the residual dependence $\deltac=\deltac(z)$ is due to the considered
cosmology.
Unfortunately, for this problem there is no analytic formulae for the
first-crossing distribution.
Nevertheless, ST02 proposed a simple analytic expression for the CMF, based on
the direct replacement of the barrier shape in their generic
approximate expression for the first-crossing distribution. This
proposal reasonably reproduces the results from numerical simulations for the
mass function of the progenitors of haloes in a given mass range
today\footnote{E.g. if we consider the conditional mass function of objects with
variance $\sigma_1$ at redshift $z_1$, given that we have a variance $\sigma_0$
at an earlier redshift $z_0$, their proposal for the re-scaling, to be inserted
in their equation for the first-crossing distribution, is $B(\sigma^2,z)
\rightarrow B(\sigma_1^2,z_1) - B(\sigma_0^2,z_0)$. }.
The ST02 proposal can not be directly applied to obtain the CMF at a fixed
redshift $z$ and for the conditioning we are considering in this paper. However,
we use here a natural extension of the ST02 proposal, which consists in the
replacement $B(\sigma^2,z) \rightarrow B(\sigma^2,z) - \delta_0$. In that case,
the explicit expression for this CMF is
\[
n_{\rm c,ST02}(m) = \Big( \frac{2}{\pi} \Big)^{1/2}
\; \frac{\meanrho}{m}
\; \frac{|T(\sigma^2|\sigma_0^2)| \sigma}{(\sigma^2 - \sigma_0^2)^{3/2}}
\Big| \frac{d\sigma}{dm} \Big|
\]
\begin{equation}
\qquad \qquad  \times \exp \Bigg\{
- \frac{[B_{\rm ec}(\sigma^2,z)-\delta_{0}]^2}{2(\sigma^2 -
\sigma_0^2)}  \Bigg\}
\label{eq:st02}
\end{equation}
where
\begin{equation}
T(\sigma^2|\sigma_0^2) = \sum_{n=0}^{5} \frac{(\sigma_0^2-\sigma^2)^n}{n!}
\frac{\partial^n [B_{\rm ec}(\sigma^2,z)-\delta_{0}]}{\partial (\sigma^2)^n}
\end{equation}
As we show below, this prescription provides better results than the EPS for
underdense regions ($\delta_0 <0$), although still shows a discrepancy with the
data from simulations, underestimating the number of haloes at low masses.

%%%%%%%%%%%
% Our method. 
\section{Extending the UMF}

Here we present our method to build the CMF of dark matter haloes.  An
important formal difference with respect to the expressions discussed in the
last section is {\it the explicit introduction of the dependence in the $q$
variable}, the distance from the center of the condition to the point at which
we evaluate the CMF.  Thus, in order to fully describe the CMF, we have to
specify three parameters, namely $Q$ (the radius of the condition), $\delta_{0}$
(the linear density within $Q$) and $q$. Thus,
\[ n_c = n_c(m, z | Q, \delta_{0}, q ). \]
Note that the quantities $Q$ and $q$ denote the Lagrangian radius; for Eulerian
radius we will use $R$ and $r$, respectively.

\subsection{The method}
The method proposed here is an extended version of the one presented
in \citet[][hereafter PBP06]{Patiri}, which was based on a set
of assumptions similar to all other previously described CMFs in the
literature. For clarity, we follow the same notation as in PBP06.
The basic idea of the method is {\it to build the CMF as an analytic
extension of any of the existing versions of the UMF}.  To this end,
we make two basic assumptions:
\begin{enumerate}
\item For those scales $m$ which are much smaller than the scale of the
condition $m(Q)$, the constrained field should behave ``locally'' as an
isotropic uniform Gaussian field with a re-scaled mean and variance (or power
spectra). This assumption is implicitly done in all the existing derivations of
the UMF (although for the average value of the field). For a discussion on its
validity, see PBP06.
\item We assume that the UMF can be derived from a ``rigid'' barrier, $B(m)$, in
the sense that if we have a crossing of this barrier, then we will have collapse
with probability $1$ (note that for this argument $B(m)$ may have some
dependence in the mass).
In the general case of ellipsoidal collapse the barrier should have certain
``width'', connected with the fact that the evolution of an ellipsoidal
perturbation is determined by three parameters (the three eigenvalues of the
deformation tensor).  Thus, for a given mass (or equivalently, a given $\sigma$)
there is no a rigid height of the barrier, but we have a probability
distribution of finite width around a certain value of $\deltal$.
However, as discussed in \citet{S01}, using an average (rigid) shape for the
barrier for treating the ellipsoidal collapse gives good results for the UMF.
\end{enumerate}

After these two assumptions, and neglecting any dependence of the shape
of the barrier which is not captured by the ratio $\delta_c/\sigma$, it
follows that obtaining the CMF just implies a vertical displacement of the
barrier (without changing its shape). This displacement is obtained as a
re-scaling in which the variables $\delta_c$ and $\sigma(m,z)$ are substituted
by their corresponding values in the local Gaussian field.
Note that a more detailed treatment of the ellipticity of the collapse
would require to compute the exact shape of the barrier after each re-scaling,
implying an integration over the distribution function for the three eigenvalues
of the deformation tensor.

% Extension of Pariri et al.
This basic idea of re-scaling the UMF is identical to the one adopted in
PBP06. However, the overall normalisation of the resulting CMF was not treated
in detail in that work, as we explain below.
Here, we extend the method by obtaining this normalisation as it is described in
Sec.~\ref{sec:normalization}, and we present the full explanation of the
approximate treatment of the normalisation used in PBP06.

%%%%%%%%%%%%
\subsection{Formalism}

The relevant statistical quantity is the conditional probability distribution,
$P(\delta_{2} | \delta_{1}, q , Q)$ for the linear density perturbation
$\delta_{2}$, on scale $Q_2$, at a distance $q$ from the center of a sphere of
radius $Q$ (the condition) with mean inner linear density fluctuation
$\delta_{1}$. For a Gaussian field, this is

\[
P(\delta_{2} | \delta_{1}, q , Q) = (2\pi)^{-1/2} 
\Big( \sigma_2^2 - \frac{\sigma_{12}^2}{\sigma_1^2} \Big)^{-1/2} \times 
\]
\begin{equation}
\qquad \exp \Big( -\frac{1}{2} \frac{ (\delta_{2} - \delta_{1} 
\frac{\sigma_{12}}{\sigma_{1}^2})^2 }{ \sigma_2^2 - 
\frac{\sigma_{12}^2}{\sigma_1^2}} \Big)
\end{equation}
where $\sigma_1^2 \equiv <\delta_1^2> = \sigma^2(Q)$, $\sigma_2^2 \equiv <\delta_2^2> =
\sigma^2(Q_2)$, and
\[
\sigma_{12} \equiv \sigma_{12}(q, Q, Q_2) =
\]
\begin{equation}
\qquad
\frac{b(z)}{2\pi^2 q} \int_{0}^{+\infty} |\delta_k|^2 
W(k Q) W(k Q_2) \sin(k q) k dk.
\end{equation}

If one compares the one-point statistic for $\delta_{2}$ with the unconstrained
case:
\begin{equation}
P(\delta_{2}) = G(\delta_2; \sigma_2) \equiv (2\pi)^{-1/2}  \sigma_2^{-1} 
\exp \Big( -\frac{1}{2} \frac{ \delta_{2}^2 }{ \sigma_2^2 } \Big)
\end{equation}
we see that the conditional case behaves in the same way, but with re-scaled
variables $\deltal'$ and $\sigma'$ given by

\begin{equation}
\deltal' =  \delta_{2} - \delta_{1} \frac{\sigma_{12}}{\sigma_{1}^2}
          \equiv  \delta_{2} - \delta_{1} D( q , Q , Q_2)
\end{equation}
and
\begin{equation}
\sigma'(m) =  \Big( \sigma_2^2 - \frac{\sigma_{12}^2}{\sigma_1^2} \Big)^{1/2}
           =  \Big( \sigma_2^2 - D(q,Q,Q_2)^2 \sigma_1^2 \Big)^{1/2}
\end{equation}
where we have introduced the function\footnote{Note that this $D(q)$ function
is independent on redshift.}
\begin{equation}
 D(q,Q,Q_2) = \frac{ \sigma_{12}(q,Q,Q_2) }{ \sigma_1^2(Q) }.
\label{eq:Dq}
\end{equation}
In Appendix~\ref{app2} we present some useful fits to this function in
several asymptotic cases.  One of the important properties that can be
used to speed up the computations with this function is that if $Q_2
\ll Q$, then $D(q,Q,Q_2)$ is practically independent of $Q_2$.

\subsection{Re-scaling the UMF}
\label{sec:rescaling}

From the previous discussion, it is natural to adopt the following
change of variables in order to re-scale locally a given UMF to obtain
the corresponding CMF:
\begin{eqnarray}
\nonumber
\delta_c  \rightarrow & \delta_c'    = \delta_c - D(q,Q,Q_2) \delta_{1} \\
\sigma^2  \rightarrow & (\sigma')^2  = \sigma^2 - D(q,Q,Q_2)^2 \sigma_1^2
\label{eq:recipe}
\end{eqnarray}
In principle, we can use any of the aforementioned expressions for the UMF to
build the CMF. However, there are some details which are important to discuss.

In the first two cases considered for the UMF (PS and ST), the UMF can
be explicitly written in terms of the variable $\nu= (\delta_c /
\sigma(m,z) )^2$.  Thus, re-scaling these mass functions just implies
the change of variables given in eq.~\ref{eq:recipe}.
However, for the WA mass function only the dependence on $\sigma^2$ is
explicit shown, so to use this UMF (or any other numerical fit) we
proceed as follows.  We shall assume that all the dependence on the
mass can be absorbed in the variable $\nu = (1.686/\sigma)^2$, and we
will re-scale this variable according to eq.~\ref{eq:recipe}. In
practise, this implies making the following change in
eq.~\ref{eq:f_wa}
\begin{equation}
\sigma^2 \rightarrow \Big( \sigma^2 - D(q,Q,Q_2)^2 \sigma_1^2 \Big)
\Big( \frac{1.686}{1.686-D(q,Q,Q_2) \delta_{1}} \Big)^2
\label{eq:recipe.v2}
\end{equation}

The BM mass function shows an additional dependence on mass apart from the
standard $\nu$ dependence, which is included in the $c(m)$ coefficient, defined
as \citep[see][]{BM2}
\begin{equation}
c( m(Q) ) \equiv - \ln [D(Q,Q,Q)]
\label{eq:cm}
\end{equation}
In order to re-scale this UMF, we use the fact that $c(m) \approx \frac{1}{2} d
\ln \sigma/ d \ln r$, so the corresponding scaling for the $c(m)$ term will be
\begin{equation}
c(m) \rightarrow c'(m) 
\approx \frac{1}{2} \frac{d\ln \sigma'}{d \ln r} = \frac{1}{4\sigma'^2} \frac{d
  \sigma'^2}{d \ln r} = c(m) \Big( \frac{\sigma}{\sigma'} \Big)^2
\label{eq:dcm}
\end{equation}
where $\sigma'$ is defined in equation~\ref{eq:recipe}.

Finally, we note that by definition, all the UMFs verify that $f(\nu) = 0$ if
$\nu<0$. This has to be explicitly taken into account because $D(q,Q,Q_2)$ can
be larger than $1$ for some values of $q$ (for example, at $q=0$), and thus,
one may find $\nu' = \delta_c'/\sigma'^2 < 0$ even if $\delta_c > \delta_{1}$.

%%%%%%%%%%%%%%%%%%%%%%%%%%%%%%%
%

%---------------
\begin{figure*}
\centering
\includegraphics[width=\columnwidth]{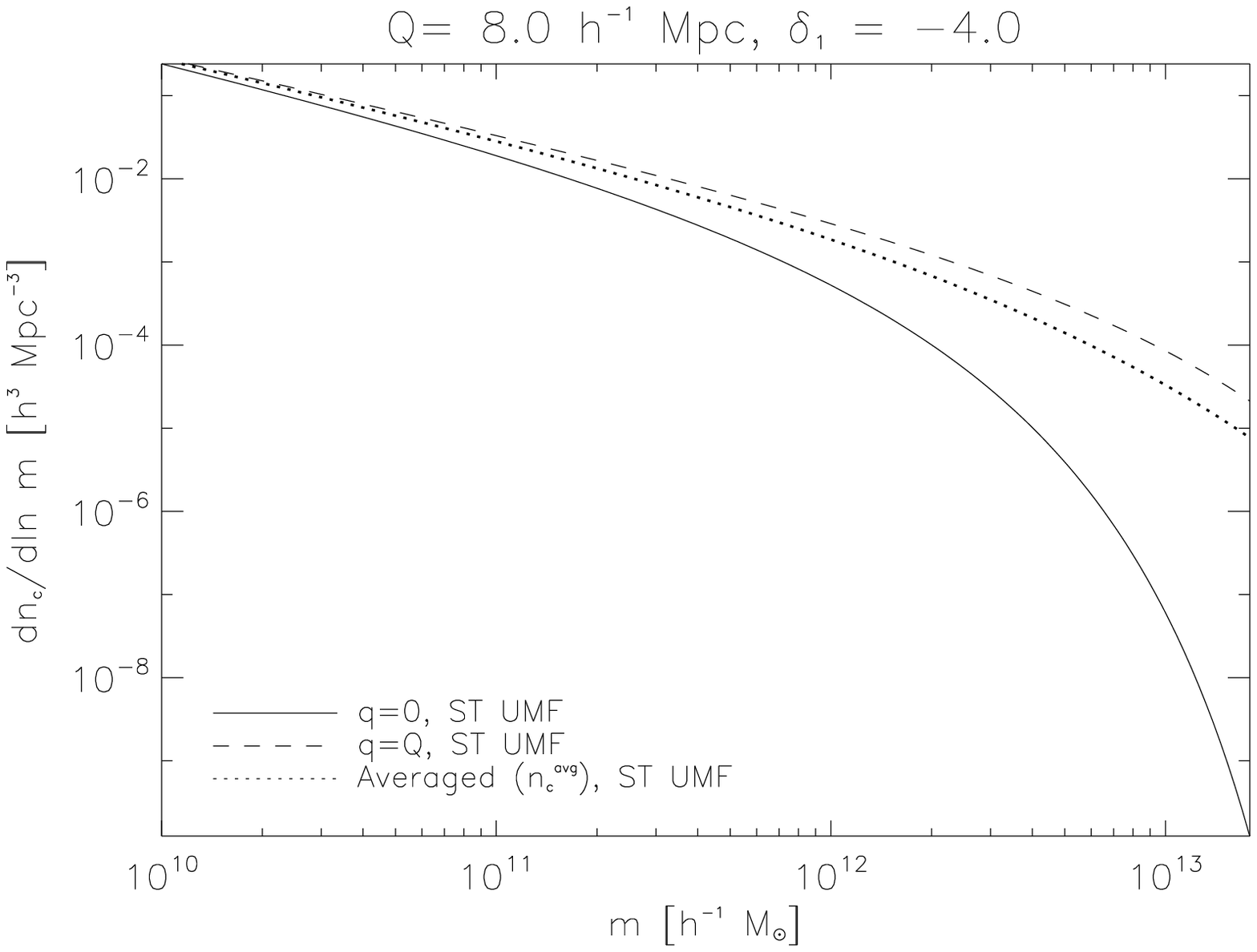}%
\includegraphics[width=\columnwidth]{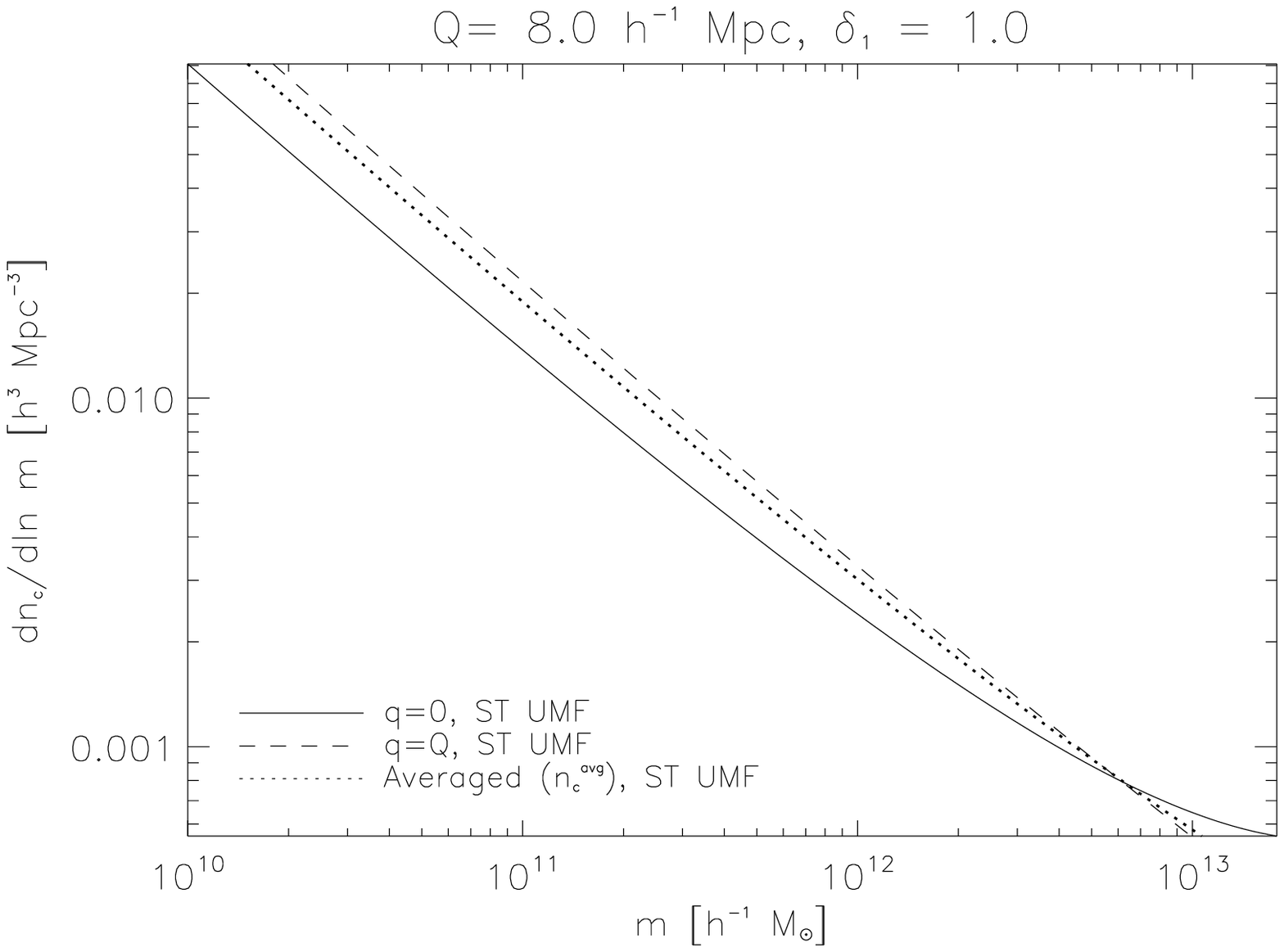}
\caption{ Radial variation of the CMF. Two examples of the CMF, $n_c(m,z| Q,
  \deltal, q)$, are shown for the case of an underdense region ($\delta_1=-4$,
  left panel) and an overdense region ($\delta_1=1.0$, right panel), for a
  condition of $Q=8$~$\runit$ (equivalent to $\approx 1.8 \times
  10^{14}$~$\munit$).  The CMF is obtained using the re-scaling presented in
  equation~\ref{eq:recipe}, and the reference UMF is taken to be the ST mass
  function.  Both panels present the two extreme cases of $q=0$ (solid line) and
  $q=Q$ (dashed line), showing the range of variation of the CMF within the
  condition. For comparison, we also present (dotted line) the average CMF over
  the condition (which is computed using equation~\ref{eq:nc_avg}). }
\label{fig:example_cmf}
\end{figure*}
%---------------

%---------------
\begin{figure*}
\centering
\includegraphics[width=\columnwidth]{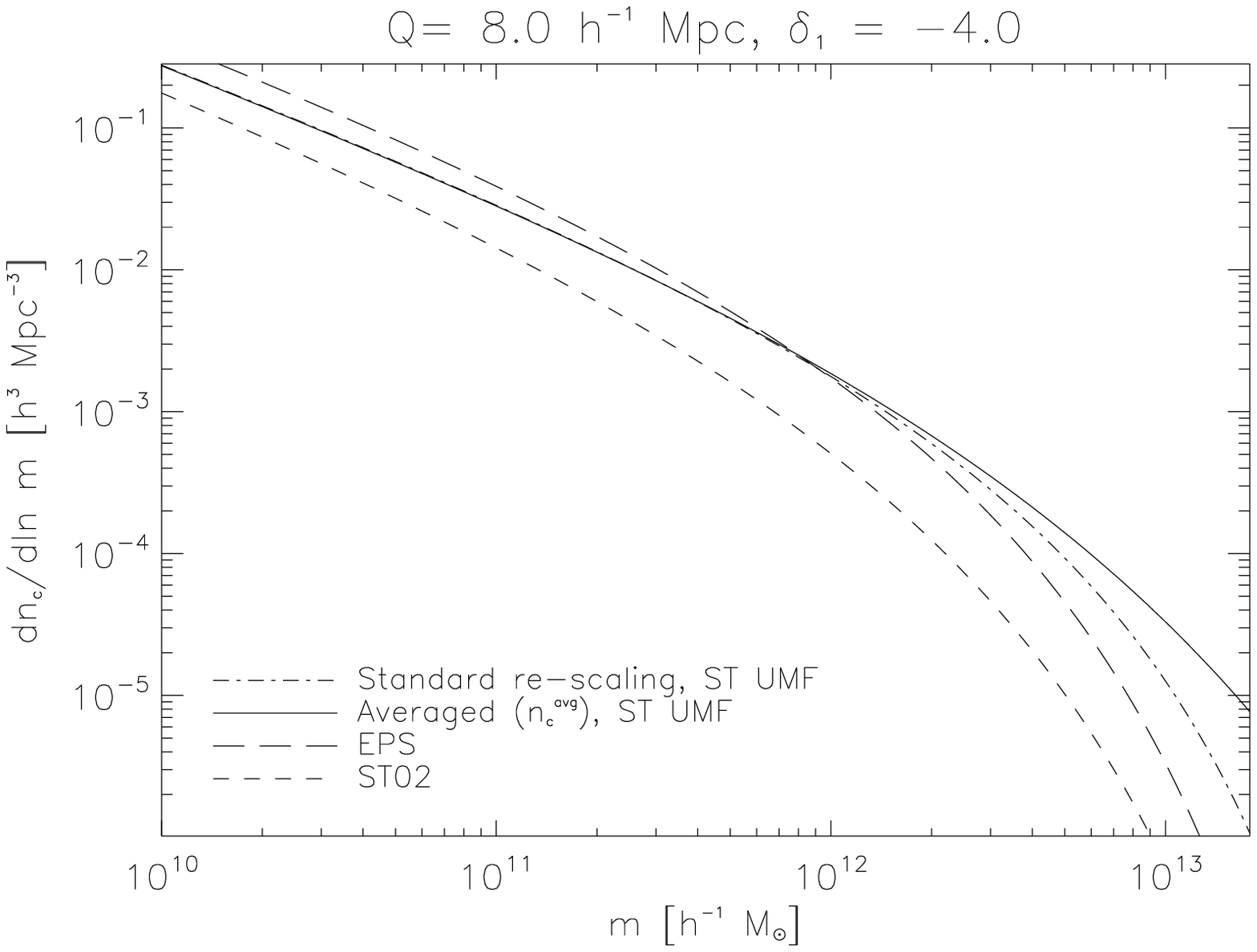}%
\includegraphics[width=\columnwidth]{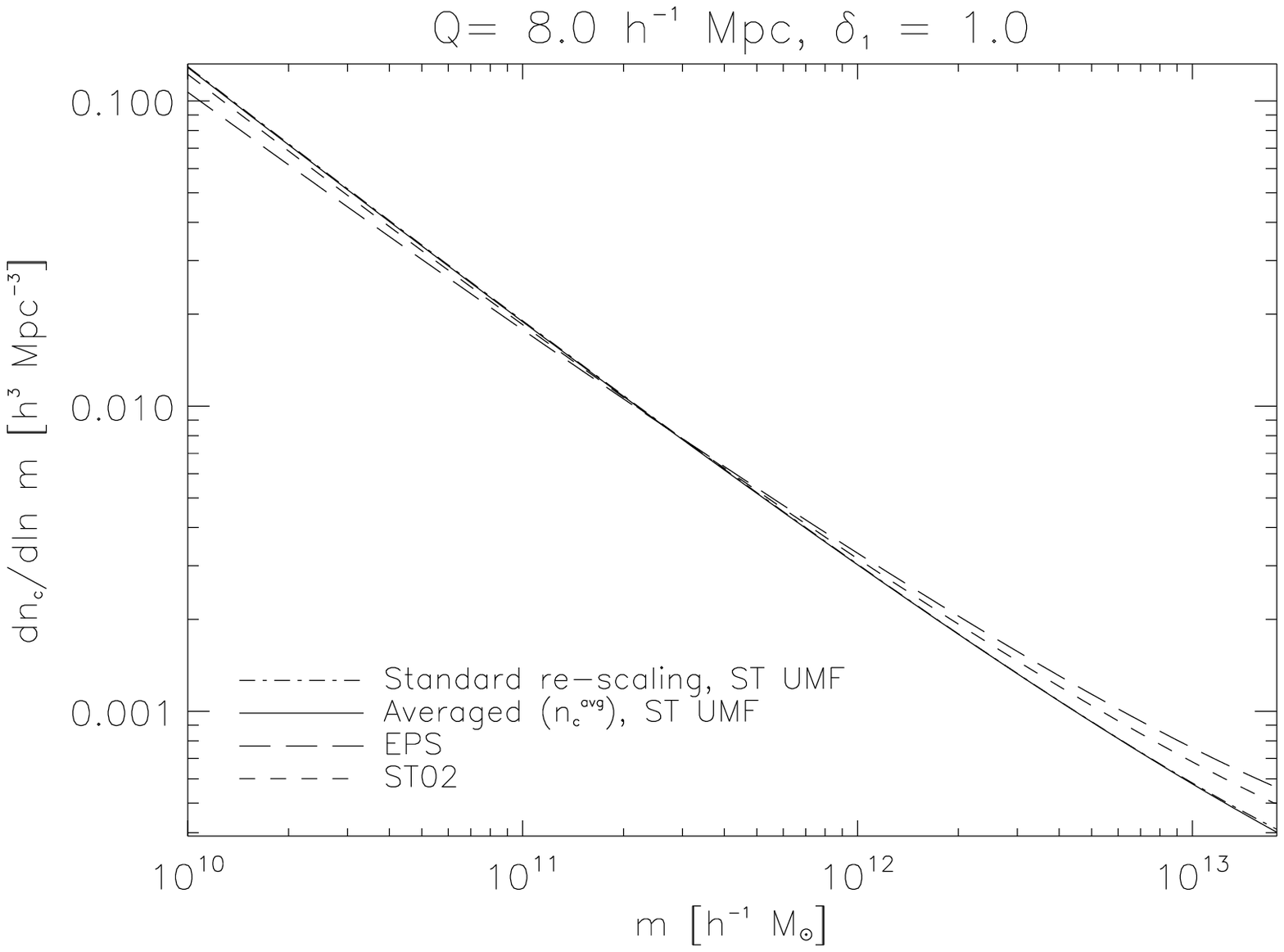}
\caption{ Comparison of the average CMF from four different prescriptions. The
 first two cases correspond to the standard rescaling (see text for details),
 and to the average CMF ($\ncavg$) computed as described in
 equation~\ref{eq:nc_avg}. Both of them are computed using the ST UMF. The other
 two prescriptions correspond to the EPS (eq.~\ref{eq:eps}) and to the ST02
 (eq.~\ref{eq:st02}).  As in Fig.~\ref{fig:example_cmf}, the left panel refers
 to an underdense region ($\delta_1=-4$), while the right panel corresponds to
 an overdense one ($\delta_1=1.0$), both of them for a condition of
 $Q=8$~$\runit$. }
\label{fig:compara_cmf}
\end{figure*}
%---------------

\subsection{Examples and comparison with other prescriptions}
\label{sec:4.4}

Fig.~\ref{fig:example_cmf} shows two particular examples of the CMF,
built in this case from the ST mass function, and using $Q=8$~$\runit$
(equivalent to $\approx 1.8 \times 10^{14}$~$\munit$). We consider the
case of an underdense region ($\delta_1=-4$), and an overdense one
($\delta_1=0.5$).
As discussed above, our prescription provides the CMF as a function of
the position within the condition, so the figure illustrates the range
of variation of $n_c$ with the radial coordinate $q$.
% (a) Case 1
For the case of $\delta_1 = -4$, the radial dependence is strong for
those masses which are a significant fraction of the total mass
of the conditioning region, being the number density
significantly smaller at the center of the condition.
% (b) Case 2
For the case of $\delta_1 = 1$, the radial dependence is important
practically at all masses. In particular, at low masses the number
density is significantly smaller at the center of the condition than
at the boundary. One can easily understand this result in this way:
near the center, $D(q) \approx 1.3$, which makes the value of the
local density very close to the threshold. In that case, most of the
mass will be collapsed in large objects (with masses of the order of a
significant fraction of $\mstar$, with $\sigma(\mstar)=1$), and low
mass object will be less abundant in number when compared with the
border.

To compare our prescription with results existing in the literature,
we need to obtain the ``volume average'' CMF over the condition,
$\ncavg$. This function is easily derived as
\begin{equation}
\ncavg (m,z | Q, \deltal) = \frac{3}{Q^3} \int_{0}^{Q} n_c(m,z | Q, \deltal, q)
q^2 dq
\label{eq:nc_avg}
\end{equation}
For such comparison, we also show in Fig.~\ref{fig:example_cmf} this average
function, which, as a consequence of the radial dependence commented above, will
be much more close in shape and amplitude to the CMF with values of $q$ in the
vicinity of $Q$.

%%%%%%%%%%%%%%%%%%%
% Fig.2
%
Fig.~\ref{fig:compara_cmf} compares these two average CMFs with other
existing prescriptions.
In particular, we first want to answer the following question: does
our averaged CMF coincides with the CMF that would be obtained using
the ``standard rescaling''? 
%(given in equation~\ref{eq:standard})?
%
In this context, we consider as ``standard re-scaling'' the one
presented in equation~(\ref{eq:standard}), {\it but applied to any of
the aforementioned UMFs and not only the PS UMF}.
The interesting result shown in Fig.~\ref{fig:compara_cmf} is that, for a given
UMF (e.g. the ST in this case), the corresponding {\it average CMF} (computed
with our formalism using equation~\ref{eq:nc_avg}), coincides with great
accuracy with that CMF obtained with the ``standard re-scaling'', but {\it only
for those masses much smaller than the condition}. However, and specially for
the case of underdense regions, when we consider larger values of the mass (in
this particular value of the linear density, for masses $\ga 10^{-2}~\mstar$),
then the ``standard'' computation differs from the exact result.

One can understand this result in the following way. Within our
approach, we are rescaling the mass function locally, according to the
local Gaussian field, and after that, we are averaging over the whole
volume. However, in the ``standard case'', the volume-averaged
re-scaling of the density field is used as an ansatz for the
prescription to rescale the UMF. In principle, these two
procedures could produce different results.
For low masses, the dependence of the mass function on the local
linear density fluctuation (whose mean value at $q$ is $\deltal
D(q,Q,Q_2)$) is close to linear, and thus the processes of
rescaling and averaging almost commute.  However, for higher masses,
non-linear terms are important and this two processes do not commute.

For comparison, Fig.~\ref{fig:compara_cmf} also presents the CMF
obtained from EPS (eq.~\ref{eq:eps}) and ST02 (eq.~\ref{eq:st02})
prescriptions. All the different mass functions give similar results
for the overdense case, while showing significant discrepancies for
the underdense case.

\section{A Normalisation condition for the CMF}
\label{sec:normalization}

In order to assess the quality (and the validity) of a given
expression for the CMF, we propose here to use a ``normalisation
condition'' that connects the CMF with the unconditional one. This
condition {\it must be satisfied by any CMF in general}, and reads
\begin{equation}
  n(m) = \int_{-\infty}^{\delta_c}  
  n_{\rm c}( m | \delta_1, Q, q=0 ) G(\delta_1; \sigma_1)
  d\delta_1, \; \;  m < m(Q)
\label{eq:norma}
\end{equation}
here $G(\delta_1; \sigma_1)$ stands for a normalised gaussian with
mean $\delta_1$ and variance $\sigma_1$, and we have omitted the
redshift dependence in both sides for simplicity. Note that, by
construction, there can not be masses larger than $m(Q)$ within the
conditioning region. It is also important to stress that in
this equation, the \emph{lagrangian CMF} is the one that has to be
used. 

Equation~\ref{eq:norma} quantifies the fact that when one integrates
the CMF for all possible values of the linear density within the
condition ($\delta_1$), then one should recover the UMF.
By definition of the CMF, it is clear that this equation is exact in
the limit $\sigma_1 \rightarrow 0$ (in that case,
$G(\delta_1;\sigma_1) \rightarrow \delta^{\rm (D)}(\delta_1)$, where
$\delta^{\rm (D)}$ is the Dirac delta-function).
%

% Validity
For finite values of $\sigma_1(Q)$, it should also be satisfied to a
good approximation for large values of $Q$ (i.e. in the linear
regime), and for those masses $m$ much smaller than $m(Q)$.
The reason for this is related with the way we have set the upper
limit of integration. In principle, one would recover the true UMF if
the integration is carried out for all possible values of $\delta_1$.
However, the CMF will not be defined for values of $\delta_1$ above
the barrier, because in that case the whole conditioning region would
have collapsed, and only masses larger than $m(Q)$ may exist (i.e. it
would be part of a larger object).
Therefore, under the assumption of a rigid barrier, the upper limit of the
integral in eq.~\ref{eq:norma} should be given by this barrier, i.e. $\deltac$
in the case of spherical collapse, or by $B_{\rm ec}(\sigma_1^2,z)$ (which
was defined in equation~\ref{eq:barrier_ec}) for the ellipsoidal collapse
barrier \citep{S01}.

As discussed above, this abrupt way of truncating the integral is
based on the assumption of rigid barrier, and it might fail for values
of $\delta_1$ close to the boundary ($\deltac$). For this
reason, and in order to normalise the CMF, we will focus on large
values of $Q$ for the computations ($\sigma_1 \la 1$). In this way,
the gaussian function in the rhs of eq.~\ref{eq:norma} is narrow, and
the details of how the truncation is done are not important,
while still eq.~\ref{eq:norma} imposes a significant
constraint on the shape of $n_{\rm c}$.

%--------------- 
\begin{figure}
\centering 
\includegraphics[width=\columnwidth]{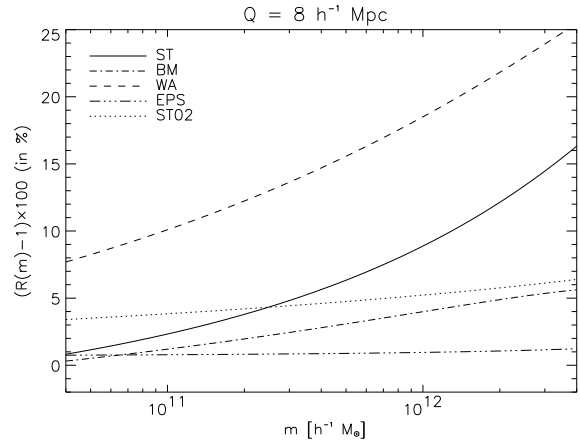}
\caption{ Relative error in the normalisation equation~\ref{eq:norma}
for the CMF. We consider the case of $R_c = 8~$Mpc/h.  This percentage
error is shown as a function of the mass $m$ for three different mass
functions: ST \citep{ST}, BM \citep{BM} and WA \citep{WA}. It can be
seen that for the ST and the WA cases, the relative error is larger
than 10\% for high masses. For comparison, it is also shown the
relative error for the case of the EPS and ST02 prescriptions. }
\label{fig:norma}
\end{figure} 
%---------------

%%%%%%%%%%%%%%%%%%%%%%%%%%%%%%%%%%%%%%%
%
Fig.~\ref{fig:norma} presents the relative difference between both sides of
Eq.~\ref{eq:norma}, defined as the ratio $R(m)$ of the right- to the left-hand
sides of that equation, i.e.
% g_int/g_ref-1.
\begin{equation}
R(m) = \frac{\int_{-\infty}^{X}
%{B_{\rm ec}(\sigma_1^2)}
%{\delta_c}  
  n_{\rm c}( m | \delta_1, Q, q=0 ) G(\delta_1; \sigma_1)
  d\delta_1}{n(m)} 
\end{equation}
where $X$ is equal to $B_{\rm ec}(\sigma_1^2,z)$ for those
cases in which the UMF includes the physics of the ellipsoidal
collapse, and $X=\deltac$ for the PS case.
For illustration, we consider three cases for the mass function
to be used with the standard re-scaling procedure, namely ST,
WA and BM, and the condition $R_c = 8\runit$ (or $1.79\times
10^{14}~\munit$). The main conclusion from that figure is that none
of the considered mass functions satisfy the normalisation condition,
showing deviations in some cases larger than 20\%, specially for high
masses.
If we repeat this computation for different values of the condition,
we find that for a given mass, the relative error becomes larger for
smaller conditions, or equivalently, for larger values of $\sigma_1$
(for example, for $R_c = 6\runit$ and the ST mass function, the
relative error becomes already of the order of 5\% for $m \approx
10^{11}~\munit$). This fact shows that the re-scaling proposed in that
case to build the CMF can not be exact.
For comparison, we also present the corresponding $R(m)$
function for the EPS and ST02 prescriptions, which can be computed
because the normalization condition (eq.~\ref{eq:norma}) also applies
to the averaged CMF.  We note that the EPS is practically normalised,
although it does not includes the physics of the ellipsoidal
collapse. On the other hand, the ST02 satisfies the normalization
condition with high accuracy (better than approx. 5 per cent),
although as it is shown below in section~\ref{sec:voids}, it does not
reproduces the numerical results in simulated voids.

%%%%%%%%%%%%%%%%%%%%%%%%%%%%%%%%%%%%%%%%%
%
\section{Extension of the method}

The incorrect normalisation of the CMFs built using the
standard re-scaling as described in the last section imply that at
least one of the assumptions made in the prescription for re-scaling
the UMF is incorrect.  Given that the hypothesis of a rigid barrier
has provided very good results for the case of UMFs, and that the
hypothesis of local isotropy should be satisfied to great accuracy,
the only remaining assumption seems to be the adopted
expression for the re-scaling.
In PBP06, the question of the normalisation was handled in a simple
manner: the re-scaled mass function was simply divided by $R(m)$. This
``first-order approximation'' procedure provided good results, but
this a posteriori renormalisation can not be entirely correct. In
principle, if one has a correct re-scaling procedure, the
normalisation condition must follow automatically.

%%%%%%%%%%%%%%%%
% our extension
%
A complete study would require a detailed derivation of the exact
shape of the ellipsoidal barrier in a conditioned environment, which
is beyond the scope of this work. Instead of that, we propose here a
simple modification of the re-scaling law in order to fulfil the
normalisation condition.
Our choice to propose a modification of eq.~\ref{eq:recipe}
is to assume that the linear density which has to be re-scaled is not
$\deltac$, but a certain parameter $\deltaeff$.
We stress that this is simply an ansatz to absorb possible errors
implicit in the re-scaling procedure given in
equations~\ref{eq:standard} and \ref{eq:recipe}, so as to render a
modified prescription which would satisfy the normalisation condition
with much higher accuracy.
This means that we should not attach much meaning to the values of
$\deltaeff$ obtained. However, as we note latter, the fact that a
particular UMF satisfies the normalisation condition with a value of
$\deltaeff$ of the order of 1.6, may be considered as a ``natural''
result and a strong point in favor of that UMF.

If the variable that has to be rescaled is $\deltaeff$ instead of
$\deltac$, then the new recipe would be
\begin{equation}
\delta_c \rightarrow \delta_c' = 
\delta_c \Bigg( 1 -  D(q,Q,Q_2) \frac{\delta_{1}}{\deltaeff} \Bigg) 
\label{eq:recipe2}
\end{equation}
or equivalently, the total re-scaling for the variance will be
\begin{equation}
\sigma^2 \rightarrow \frac{ \sigma^2 - D(q,Q,Q_2)^2 \sigma_1^2 }{ ( 1 -
D(q,Q,Q_2) \delta_{1}/ \deltaeff )^2}
\label{eq:recipe2.v2}
\end{equation}

%----------------------------------------------
\begin{table}
\caption{Values for the $\deltaeff$ parameter that satisfy the normalisation
condition (Eq.~\ref{eq:norma}) for the different UMFs considered in this
paper. These values have been obtained from a condition of 6~$\runit$.  For each
one of the UMF presented in the first column, the second column shows the
interval in which the error in the normalisation condition is smaller than 5\%,
at the largest considered mass ($Q_{\rm void}/30$).  Third column shows the
``optimum'' $\deltaeff$ which we finally adopt, obtained as the value that gives
the smallest percentage error in the normalisation condition at the highest
considered mass ($Q_{\rm void}/30$). }
\label{tab:alphas}
\centering
\begin{tabular}{@{}lcc}
\hline
\hline
UMF & $\deltaeff$         & $\deltaeff$ \\
    & $[{\rm Err} < 5\%]$ &  (Adopted) \\ 
\hline
PS  & $[1.31, 1.53]$ & 1.42 \\
ST  & $[1.20, 1.42]$ & 1.25 \\
WA  & $[1.02, 1.19]$ & 1.10 \\
BM  & $[1.37, 1.64]$ & 1.50 \\
\hline
\hline
\end{tabular}
\end{table}
%----------------------------------------------

%---------------
\begin{figure*}
\centering
\includegraphics[width=\columnwidth]{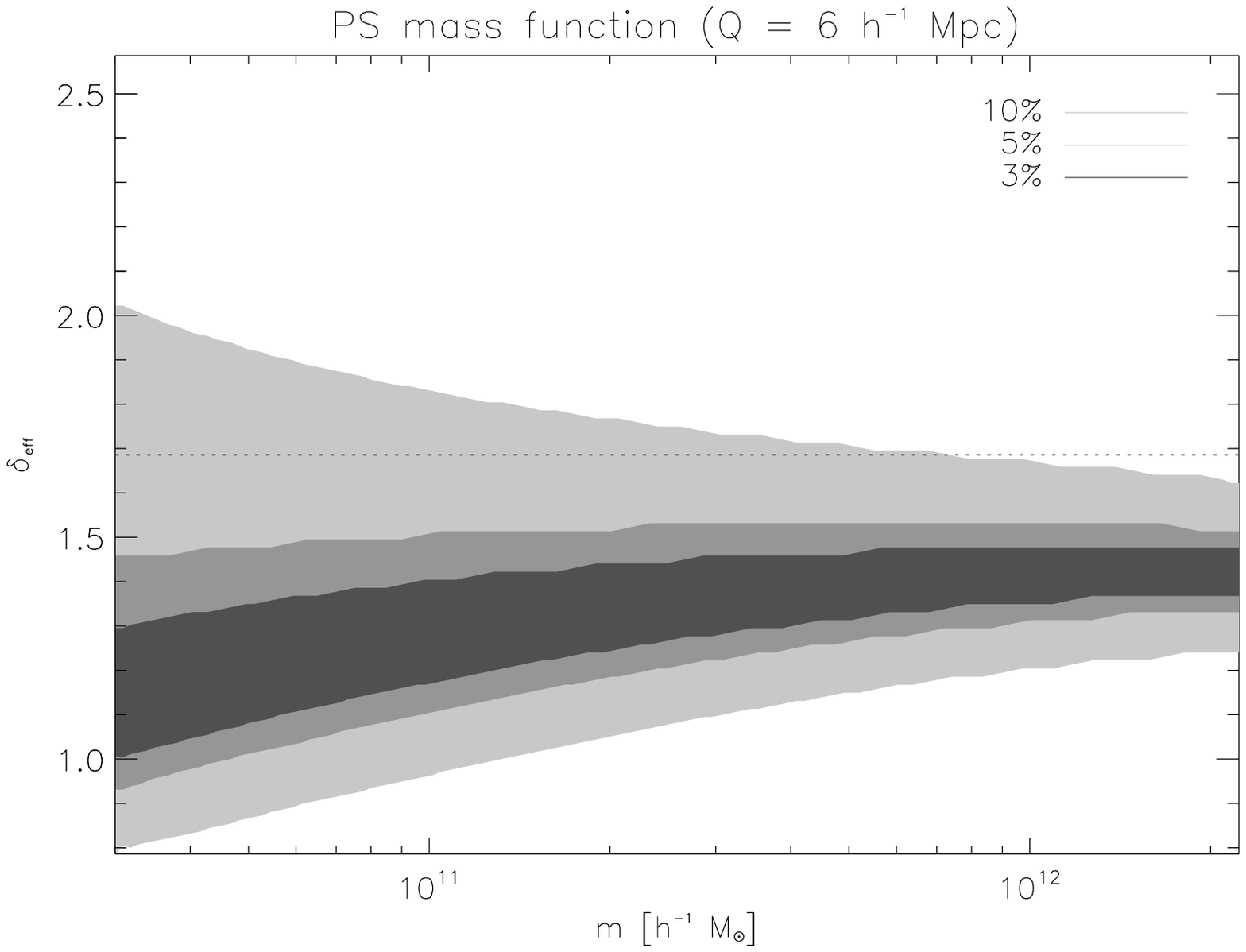}%
\includegraphics[width=\columnwidth]{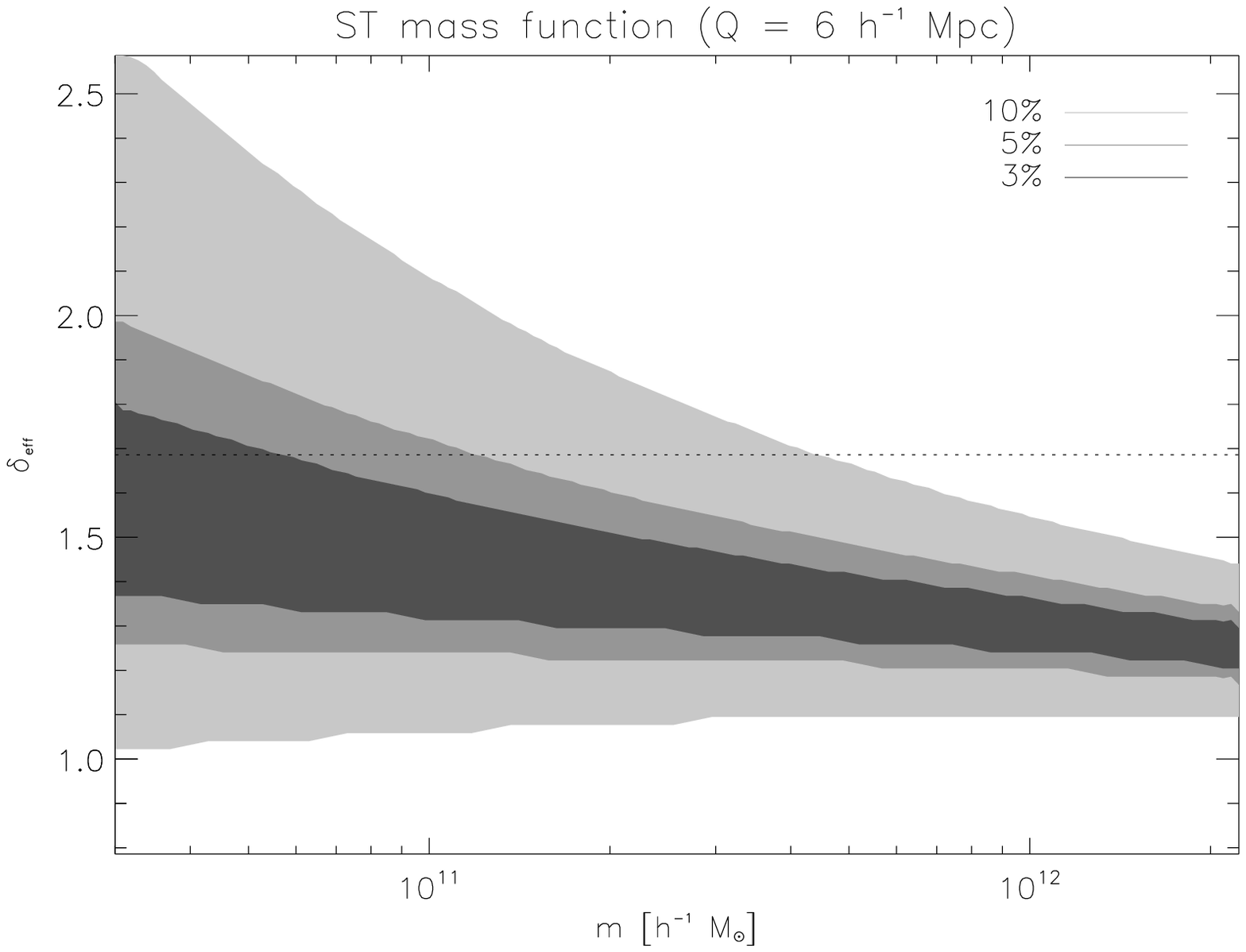}
\includegraphics[width=\columnwidth]{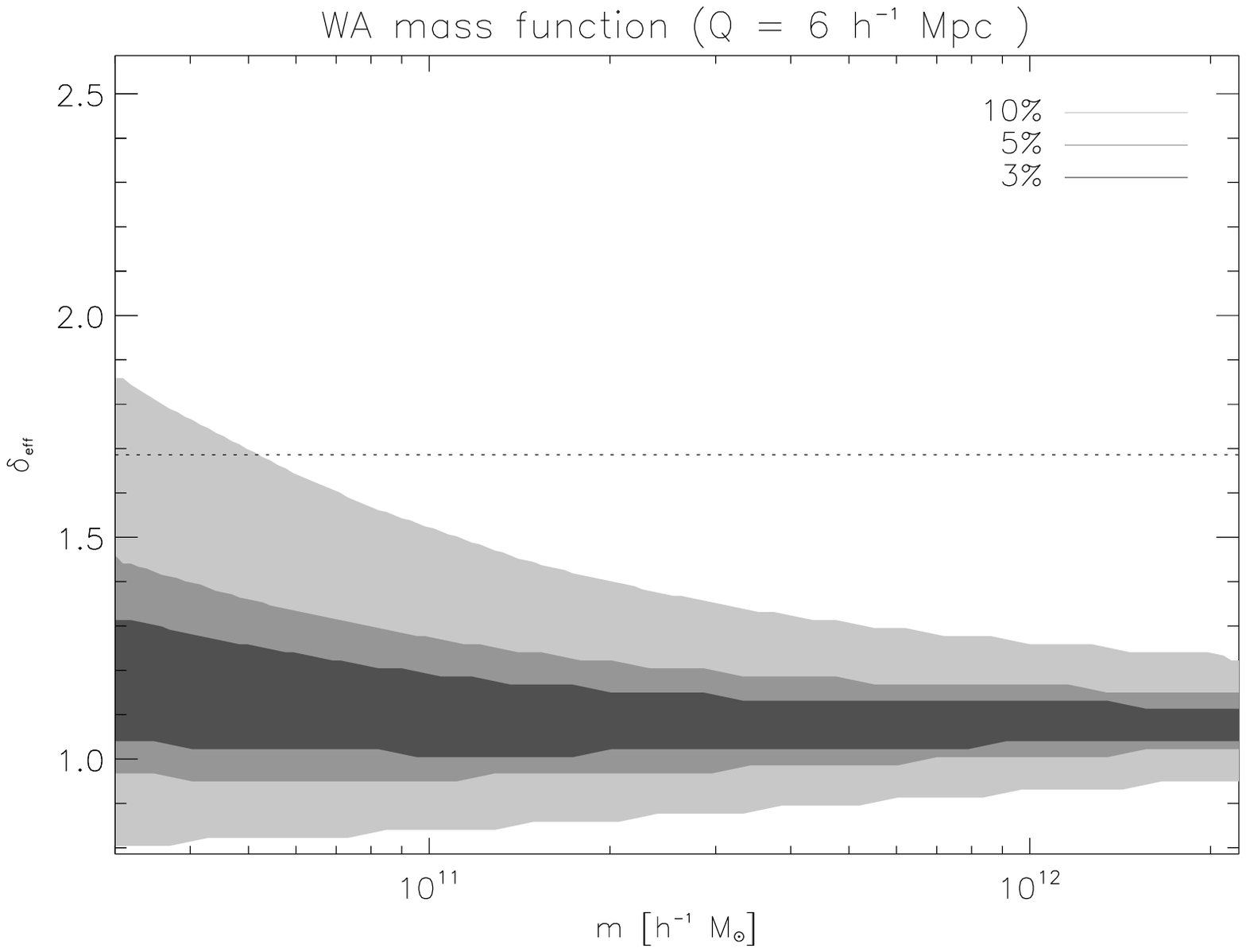}%
\includegraphics[width=\columnwidth]{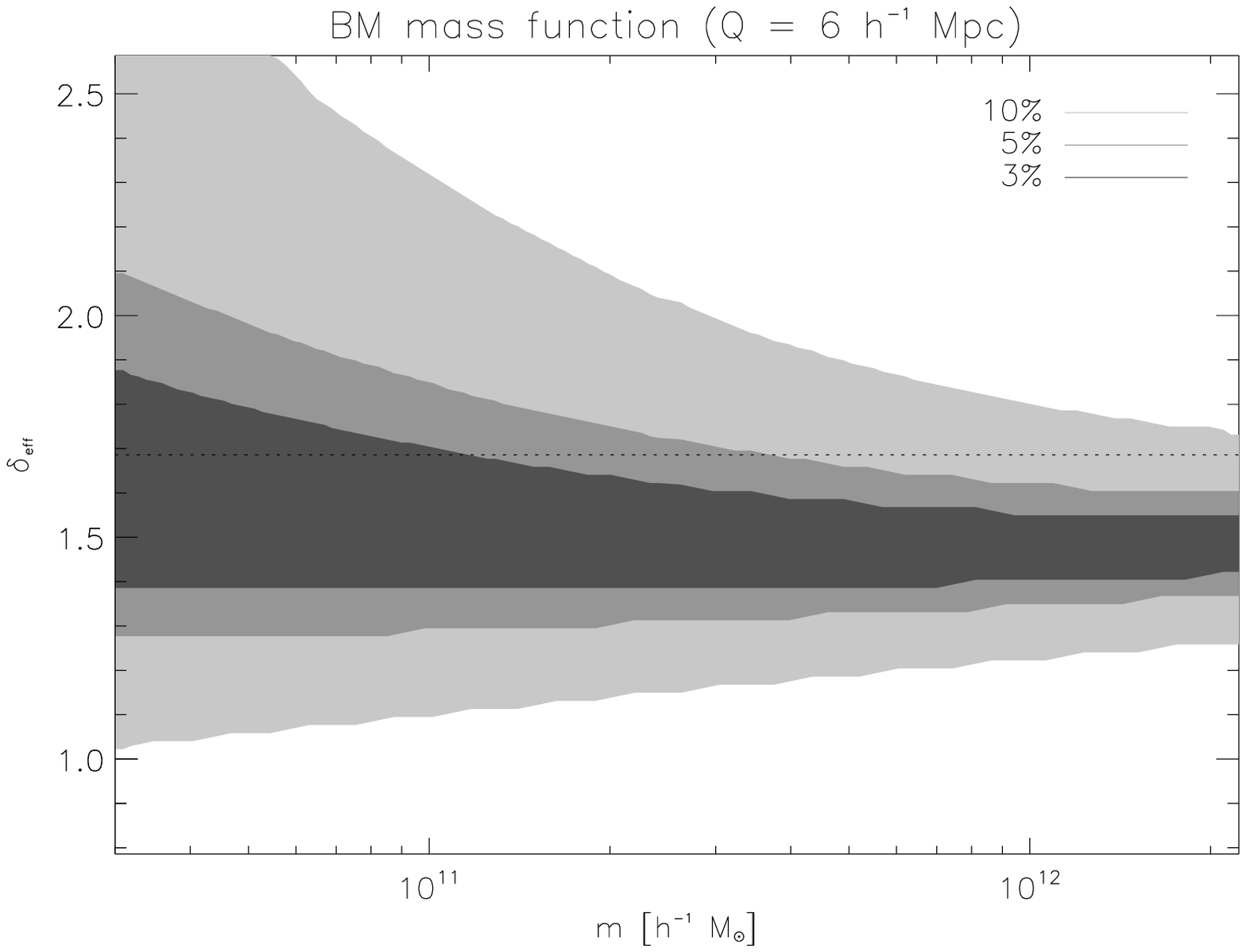}
\caption{ Ranges of the $\deltaeff$ parameter which make the normalisation
condition to be fulfilled with certain precision (10, 5 and 3 per cent,
respectively), as a function of the mass. We consider four different
prescriptions for the mass function: PS (upper left), ST (upper right), WA
(lower left) and BM (lower right). All computations have been done for the case
of $Q = 6~\runit$. }
\label{fig:deltaeffs}
\end{figure*}
%---------------

%
\subsection{Calibrating the $\deltaeff$ parameter}
Using this new prescription (equations~\ref{eq:recipe2} and
\ref{eq:recipe2.v2}) to build the CMF, we can solve the normalisation
condition (equation~\ref{eq:norma}) to find a solution for
$\deltaeff$, which in principle would be a function of mass.
Given that the different expressions of the UMF are not exact, but
they have numerical uncertainties (typically of the order of few
percent), we decided to find, for each mass, the range of $\deltaeff$
values that fulfil the normalisation condition within a certain
accuracy. Then, we look for a single value of $\deltaeff$ which
satisfies eq.~\ref{eq:norma} with high accuracy for all masses.

Figure~\ref{fig:deltaeffs} presents the allowed region for $\deltaeff$
as a function of mass, which fulfil the normalisation condition for a
certain UMF with an accuracy better than 10, 5 and 3 per cent,
respectively.  For this figure, the case of $Q=6~\runit$ was
considered, and the upper limit of the integral in
equation~\ref{eq:norma} was taken to be the barrier, $B_{\rm
ec}(\sigma^2_1,z)$.
As one would expect, for a given UMF the strongest constraint on $\deltaeff$ is
obtained for high masses, because at those masses the differences between the
CMF and the UMF become more important.
If our ansatz were the correct scaling, then one should be able to find a value
of $\deltaeff$ for which the error in the normalisation would be zero for all
masses. This is not the case, for example, for the PS UMF, where the ``allowed
region'' changes with mass. The reason for this could be that this is the only
UMF which does not take into account the physics of the ellipsoidal collapse.
However, the important point is that in all the other cases (ST, BM and WA) it
is possible to select a value for $\deltaeff$ which makes the error in the
normalisation condition to be smaller than $\sim 5-10$\% for all masses.

\begin{figure}
\centering
\includegraphics[width=\columnwidth]{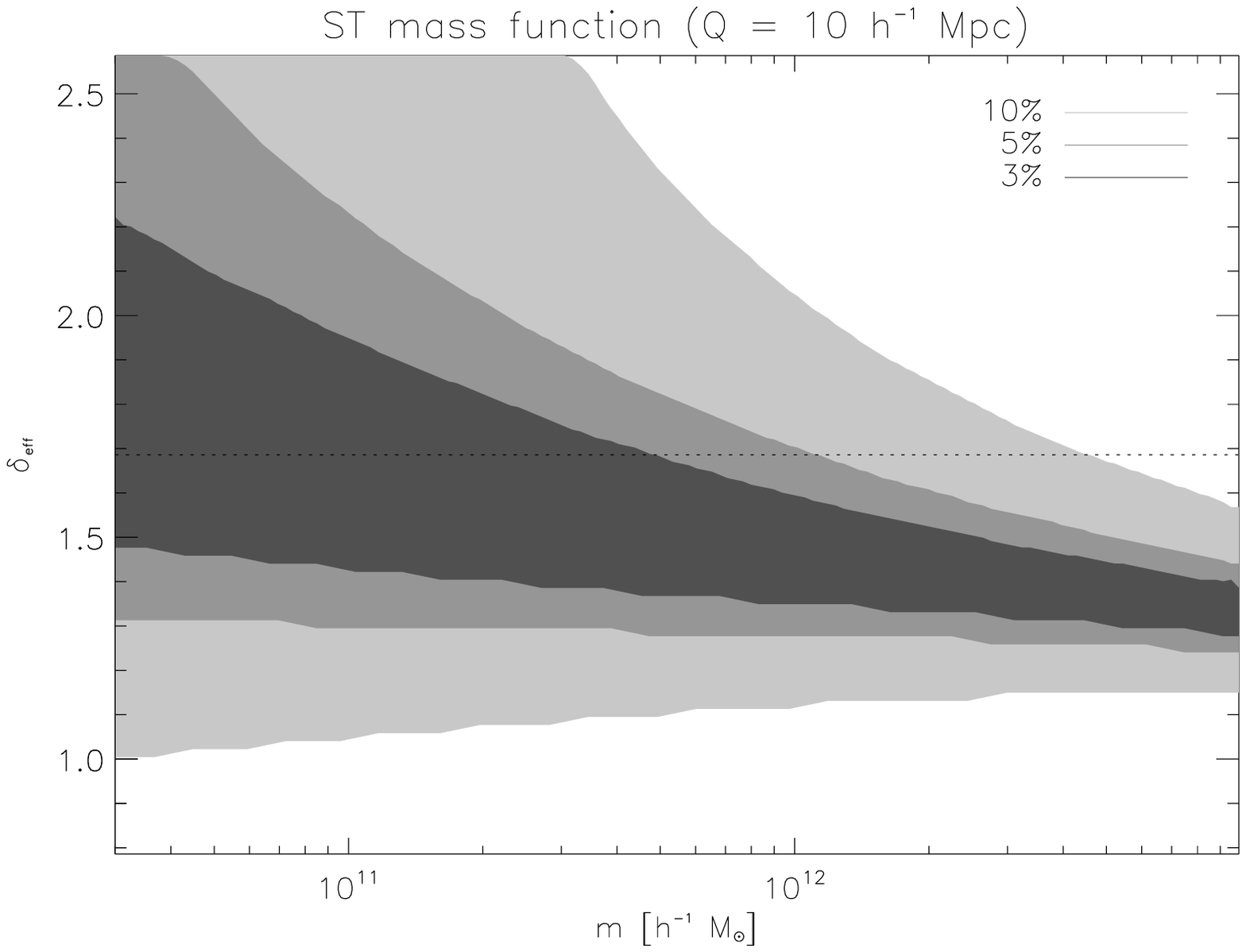}
\caption{ Same as figure~\ref{fig:deltaeffs}, but for the case of $Q =
  10~\runit$. Although the allowed region at the level of 5\% is compatible with
  the one obtained for $Q = 8~\runit$ in Fig.~\ref{fig:deltaeffs}, in this case
  it is much larger, and thus less restrictive if one is interested in
  constraining the value of $\deltaeff$. }
\label{fig:deltaeffs2}
\end{figure}

%
% Dependence with the size of the condition
Another important issue is the dependence of the $\deltaeff$ with the size of
the condition, $Q$. If our ansatz were the correct scaling, then this parameter
should not depend on $Q$.
We have checked that the $\deltaeff$ values do not show a strong dependence on
the chosen size for the condition, in the sense that the range of $\deltaeff$
which fulfils the normalisation equation with accuracy better than 5\% is
always overlapping when changing the size of the condition between $5~\runit$
and $15~\runit$. 
Indeed, choosing a value too high for the condition does not provide a good
constraint on $\deltaeff$, because in that limit the conditioning fluctuation is
very small and it is not imposing strong constraints on the exact shape of the
CMF. We illustrate this fact in Figure~\ref{fig:deltaeffs2}, where we show the
allowed regions for the case of the ST UMF, but considering now the condition
$Q=10~\runit$.
For this reason, we decided to adopt as reference values for this paper those
obtained with the case $Q=6~\runit$.  

%
% Final values for this paper
%
The adopted values of $\deltaeff$ for the computations in the rest of this paper
are presented in Table~\ref{tab:alphas}.  Our criteria was to take as
``adopted'' $\deltaeff$ value the one giving the lowest relative error at the
highest considered mass in each case, which in our computations was taken to be
$m(Q)/30$. However, we note that as regards to the normalisation condition, any
of the values quoted in the 5 per cent range could be adopted. 
%

%---------------
%---------------

\subsection{Examples of the CMF with the $\deltaeff$ parameter included}
\begin{figure*}
\centering
\includegraphics[width=\columnwidth]{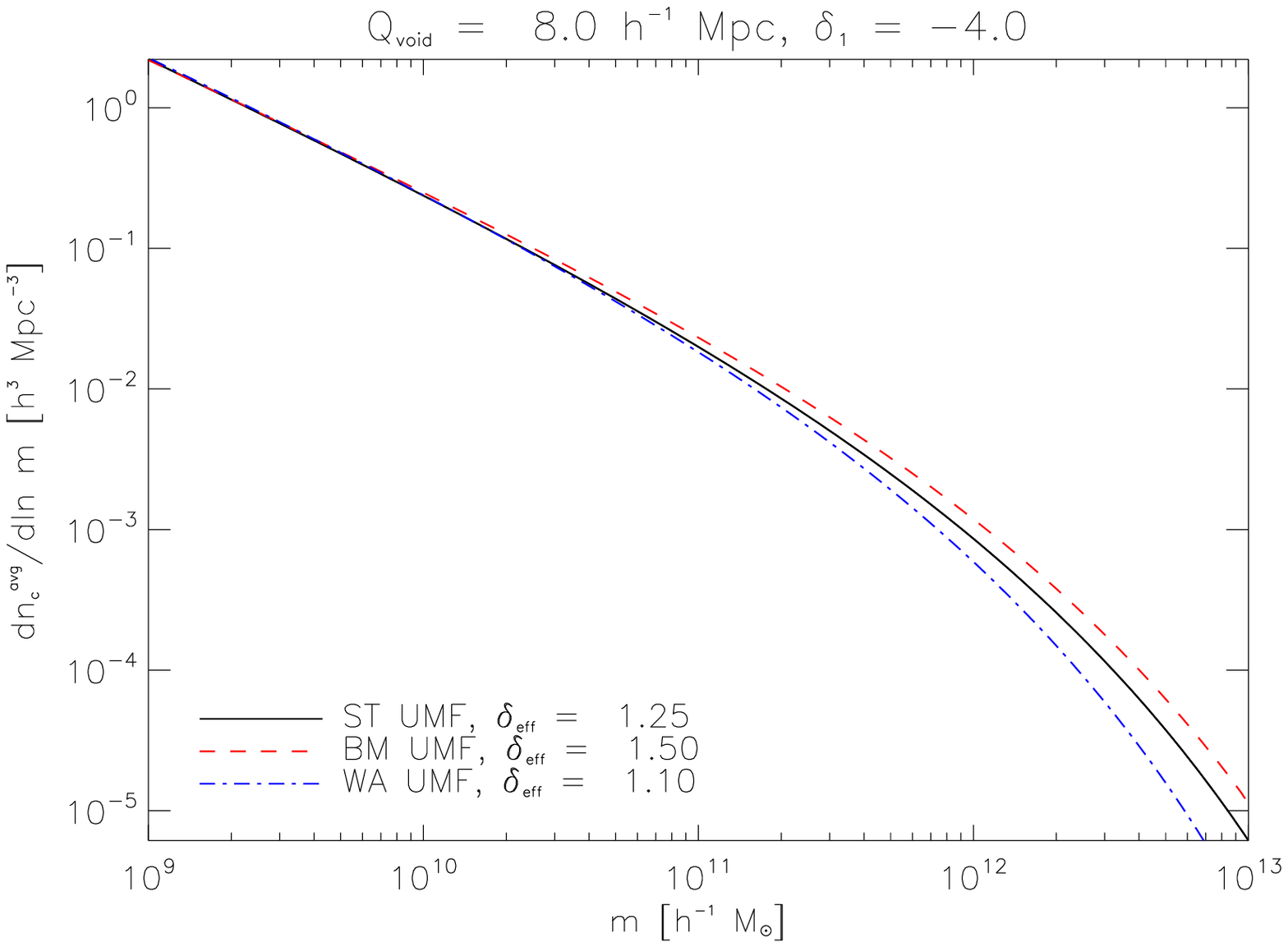}%
\includegraphics[width=\columnwidth]{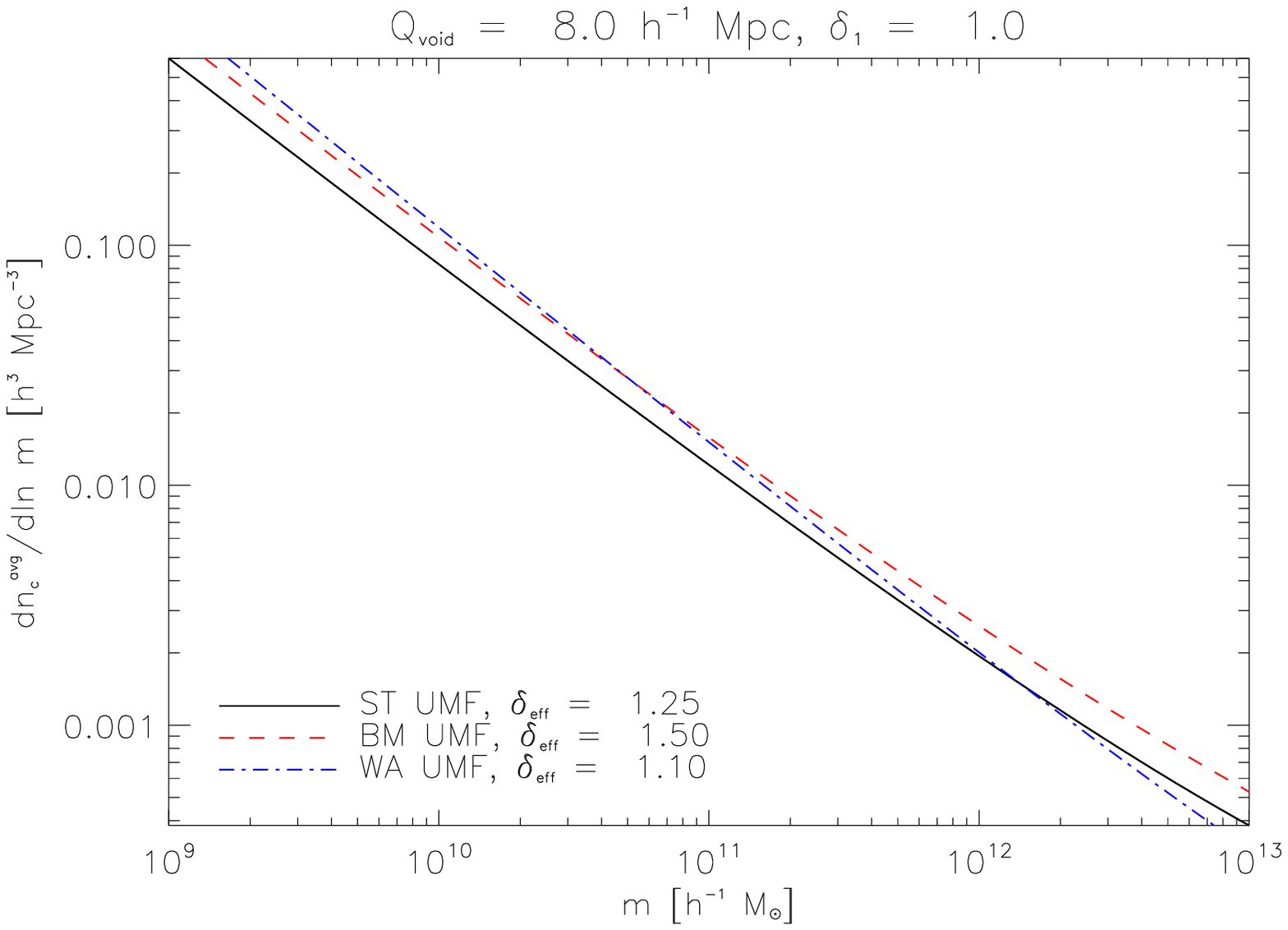}
\caption{ Comparison of the average CMF ($\ncavg$) obtained within our formalism
  (i.e. using the scaling given in eqs.~\ref{eq:recipe2} and
  \ref{eq:recipe2.v2}), but using different UMFs. We consider the ST, BM and WA
  UMFs, with their corresponding $\deltaeff$ value taken from
  Table~\ref{tab:alphas}, and for the same particular cases discussed before in
  the paper ($Q=8$~$\runit$, and either $\delta_1=-4$ or $\delta_1=1.0$).  }
\label{fig:several_umfs}
\end{figure*}

% Values of deltaeff are different
The values of $\deltaeff$ obtained for different fits to the UMF are somewhat
different, but for each particular case, they lead to similar shapes for the
CMF. We illustrate this in figure~\ref{fig:several_umfs}, where we present the
average CMF for the two cases considered before ($Q=8~\runit$, with $\delta_1 =
-4$ or $\delta_1=1$), and using several UMFs with their corresponding
``optimum'' $\deltaeff$ values. 
Given that the normalisation condition is fulfilled by all the different CMFs
with a precision of $\sim 5$\%, one would expect differences of this order
between them. However, this is not the case.  For the low-density case
($\delta_1=-4$), there are significant differences at high masses, although the
agreement at low masses is very good.  And for the high density one
($\delta_1=1$), the discrepancies are larger than that even at low masses.
%

%---------------
%---------------
%
%---------------
\begin{figure*}
\centering
\includegraphics[width=\columnwidth]{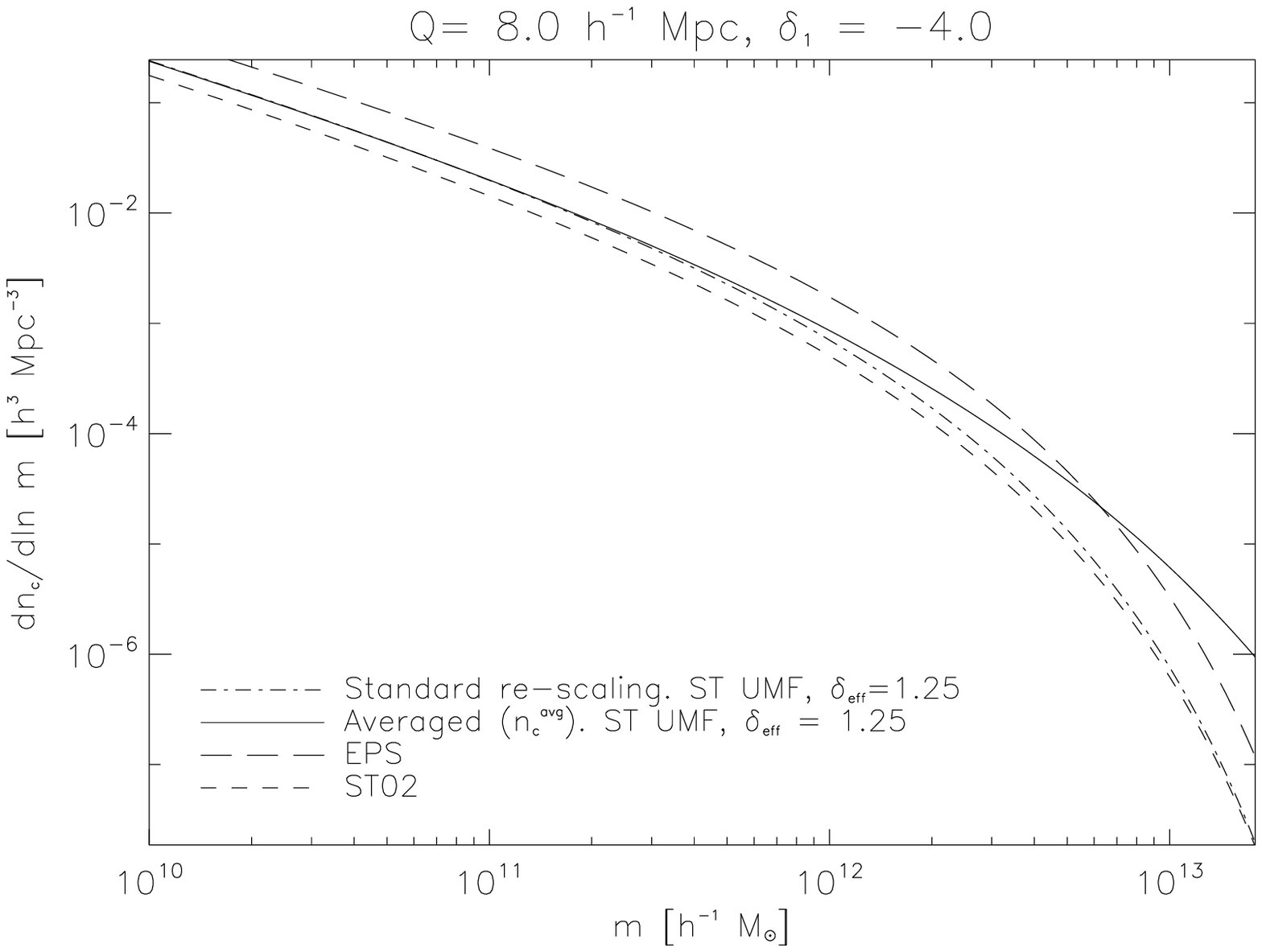}%
\includegraphics[width=\columnwidth]{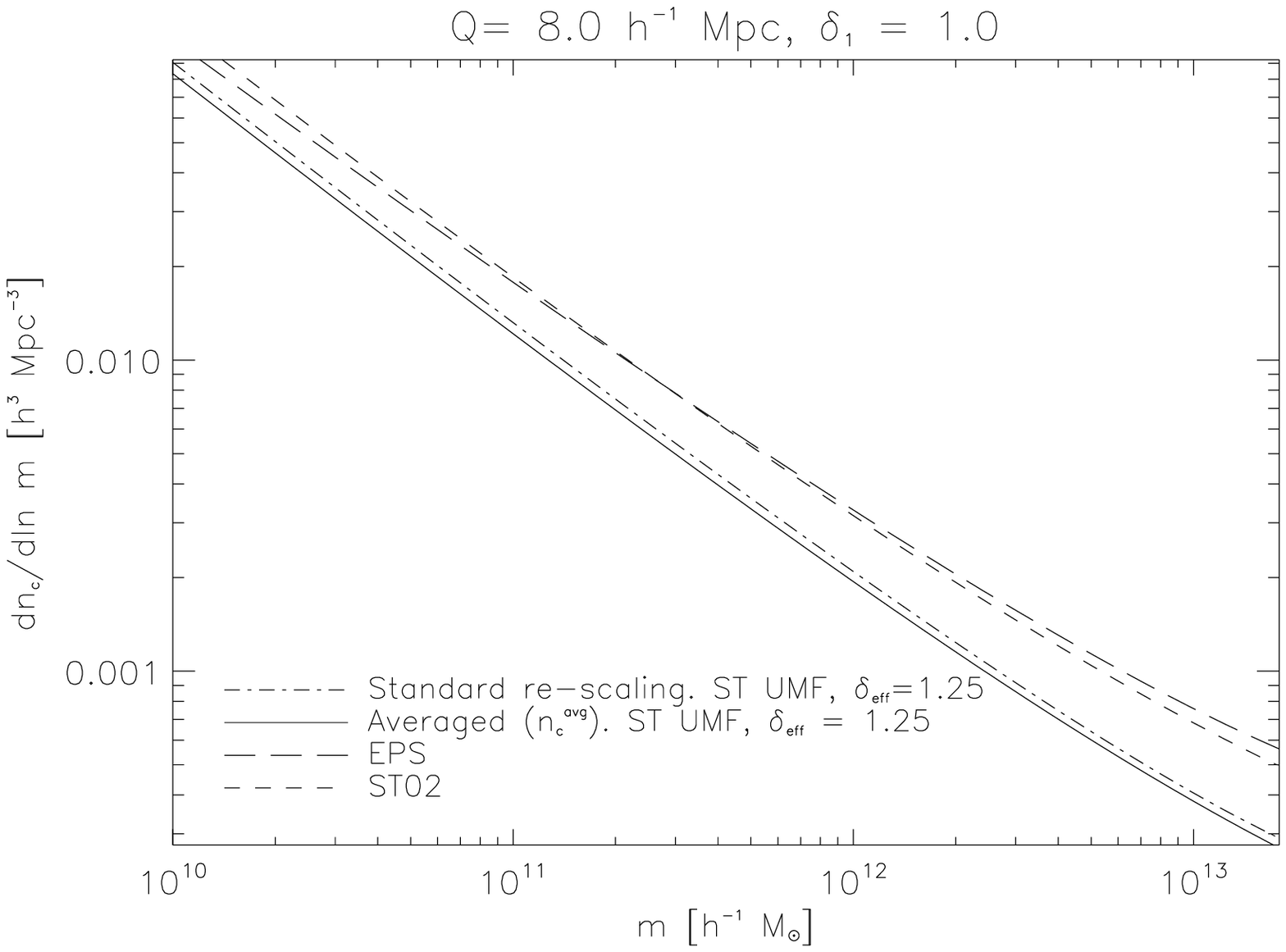}
\caption{ Comparison of the average CMF from four different prescriptions. This
  figure is equivalent to Fig.~\ref{fig:compara_cmf}, but now we use the optimum
  $\deltaeff$ instead of $\deltac$ when plotting the CMF. The first two
  prescriptions correspond to the standard re-scaling (eq.~\ref{eq:standard}),
  and to the method proposed in this work (eq.~\ref{eq:recipe2}), both using the
  ST UMF, so $\deltaeff = 1.25$ according to Table~\ref{tab:alphas}. The EPS and
  the ST02 CMFs are also shown for comparison.  As in
  Fig.~\ref{fig:compara_cmf}, we consider a condition of $Q=8$~$\runit$, and
  $\delta_1=-4$ (left panel) or $\delta_1=1.0$ (right panel). }
\label{fig:compara_cmf2}
\end{figure*}
%---------------

To conclude this subsection, we revisit again the discussion about the
difference between our averaged CMF, and the CMF obtained with the standard
(average) re-scaling of the UMF, but now using $\deltaeff$ instead of $\deltac$
as the variable to re-scale.  We have already shown in Sect.~\ref{sec:4.4} that
for the case of $\deltaeff=\deltac$, the standard re-scaling provides good
results for low masses (see Fig.~\ref{fig:compara_cmf}). As a example, we now
consider in Figure~\ref{fig:compara_cmf2} the case of $\deltaeff=1.25$ for the
ST UMF.
The result for the under-dense region ($\delta_1=-4$) is qualitatively similar to
the one obtained before for the case of $\deltaeff=\deltac$. Indeed, the
discrepancies between the averaged CMF and the CMF with the average standard
re-scaling appear approximately at the same mass scale ($m\sim
10^{-2}~\mstar$). Thus, the ``average'' re-scaling still provides good results
for those masses much smaller than the size of the conditioning region.
However, this is not the case for the over-dense ($\delta_1=1$) region, in which
even at low masses, there is a discrepancy with the exact result.  The reason
for that is that in this case, the region near the center has surpassed the
critical value of $\deltaeff$ (note that $\deltaeff=1.25$ for the ST UMF, and
that $D(q=0)\approx 1.3$), so in practice that region is not contributing to the
average. As a consequence, the CMF built with standard re-scaling slightly
overestimates the number of low-mass objects.

Finally, it is interesting to compare these results with the EPS and ST02
prescriptions, which are also presented in Fig.~\ref{fig:compara_cmf2}.  For the
under-dense region ($\delta_1=-4$), our final version of the CMF is now in closer
agreement with the ST02, although there is still a systematic offset at low
masses between the two prescriptions of the order of 30-40\%. The discrepancy
with the EPS is much larger. As can be seen from the numerical results presented
in the next section, this later prescription significantly over-estimates the
number of haloes at low masses.
For the over-dense region ($\delta_1=1$), both EPS and ST02 are systematically
above our final result.

%%%%%%%%%%%%%%%%%%%%%%%%%%%%%
%%%%%%%%%%%%%%%%%%%%%%%%%%%%%
%
\section{The conditional mass function in voids}
\label{sec:voids}

One of our main motivations for the development of an accurate prediction for
the CMF is the theoretical study of the statistic of voids.
PBP06 developed a general analytical procedure for computing the number density
of voids with radius above a given value, which was applied successfully to the
description of the statistics of voids found in numerical simulations.

Their formalism is based on a detailed study of the number density of
non-overlapping empty spheres with radius $r$. PBP06 provided an analytical
expression relating this quantity with the void probability function (VPF),
$P_0(r)$ (i.e. the probability that a randomly placed sphere of radius $r$ is
empty).
$P_0(r)$ can be obtained from two basic ingredients, namely the probability
distribution function for the values of the density contrast, $\delta$, within a
randomly chosen sphere of radius $r$; and $\delta_N(\delta_{l})$, the mean
fractional fluctuation within $r$ of the number density of the objects defining
the void as a function of the linear fractional density fluctuation within
$r$. In detail, the final expression is
\begin{equation}
P_0(r) = \int_{-\infty}^{+\infty} 
{\rm e}^{-\overline{n}V[1+\delta_N(\deltal)]} P(\delta_l|r) d\deltal
\end{equation}

In this section we will focus in the $\delta_N(\delta_{l})$ function, because it
encodes all the information about the CMF. This function can be written as
\begin{equation}
1 + \delta_N(m,Q,\deltal) = [1 + \delta_{\rm ns}(m,Q,\deltal) ]
[1 + \delta(\deltal)]
\end{equation}
where $\delta_{\rm ns}$ accounts for the ``statistical fluctuation''
(i.e. the clustering of the protohaloes in the initial conditions
before they move with mass), and $\delta(\deltal)$ is the actual
fractional mass density fluctuation as a function of the linear value.
As shown in PBP06, the $\delta_{\rm ns}$ term can be related to the CMF in 
the following way
\begin{equation}
1 + \delta_{\rm ns}(m,Q,\deltal) = \frac{1}{N(m)} \Bigg[ \frac{3}{Q^3}
\int_{0}^{Q} N_c(m | Q, \deltal, q ) q^2 dq \Bigg]
\label{eq:deltans}
\end{equation}
where we have introduced the number density of collapsed objects with masses
above $m$ for the conditional case, which is given by
\begin{equation}
N_c(m|Q,\deltal,q) = \int_{m}^{+\infty} n_c(m'|Q,\deltal,q) dm'
\end{equation}

%
% Results
We can use now our proposed formalism to produce accurate computations
for the $\delta_{\rm ns}(m,Q,\delta_l)$ function which can be used
when applying the aforementioned formalism. As proposed in PBP06,
equation~\ref{eq:deltans} can be well-described with a function of the shape
\begin{equation}
1 + \delta_{\rm ns}(m,Q,\deltal) = A(m,Q) {\rm e}^{-b(m,Q) \deltal^2}
\label{eq:Ab}
\end{equation}
We have checked that this fitting formulae provides a reasonable fit for those
masses $m \ga 3\times 10^{-3}~ \mstar$ (with $\sigma(\mstar)=1$).  Detailed
computations of the $A(m,Q)$ and $b(m,Q)$ functions for the cosmology
considered in this paper ($\sigma_8=0.9$ and $\Gamma=0.21$) are given in
appendix~\ref{app3}. The dependence on cosmology of the different coefficients
of the fits is presented in appendix~\ref{app4}.

% COmparision with Gottlober et al. 
Finally, as an illustration of our method, we compare the CMFs of haloes within
voids with the results of numerical simulations presented in \cite{Gottlober}.
Figure~5 in that paper presents the mass function of haloes in five simulated
voids for a cosmology with $\Omega_m = 0.3$. Two of them correspond to voids
with radius $R_{\rm void} = 10\runit$ and mean density (in units of the critical
density) of $0.03$ (i.e. the density contrast is $\delta=0.03/0.3-1=-0.900$);
while the other three correspond to voids of radius $R_{\rm void} = 8\runit$ and
mean density $0.04$ (or equivalently, $\delta=-0.867$).
In order to convert the mean density values ($\delta$) into linear density ones,
we use the expression given by \citet{S02},  
\[
\deltal (\delta ) = \frac{\deltac}{1.68647} \Bigg[ 1.68647 -
  \frac{1.35}{(1+\delta)^{2/3}} - \frac{1.12431}{(1+\delta)^{1/2}} 
\]
\begin{equation}
  \qquad  \qquad \qquad + \frac{0.78785}{(1+\delta)^{0.58661}}  \Bigg] 
\end{equation}
For the two considered voids, we obtain $\deltal = -5.092$ and $\deltal =
-3.995$, respectively.

\begin{figure*}
\centering
\includegraphics[width=\columnwidth]{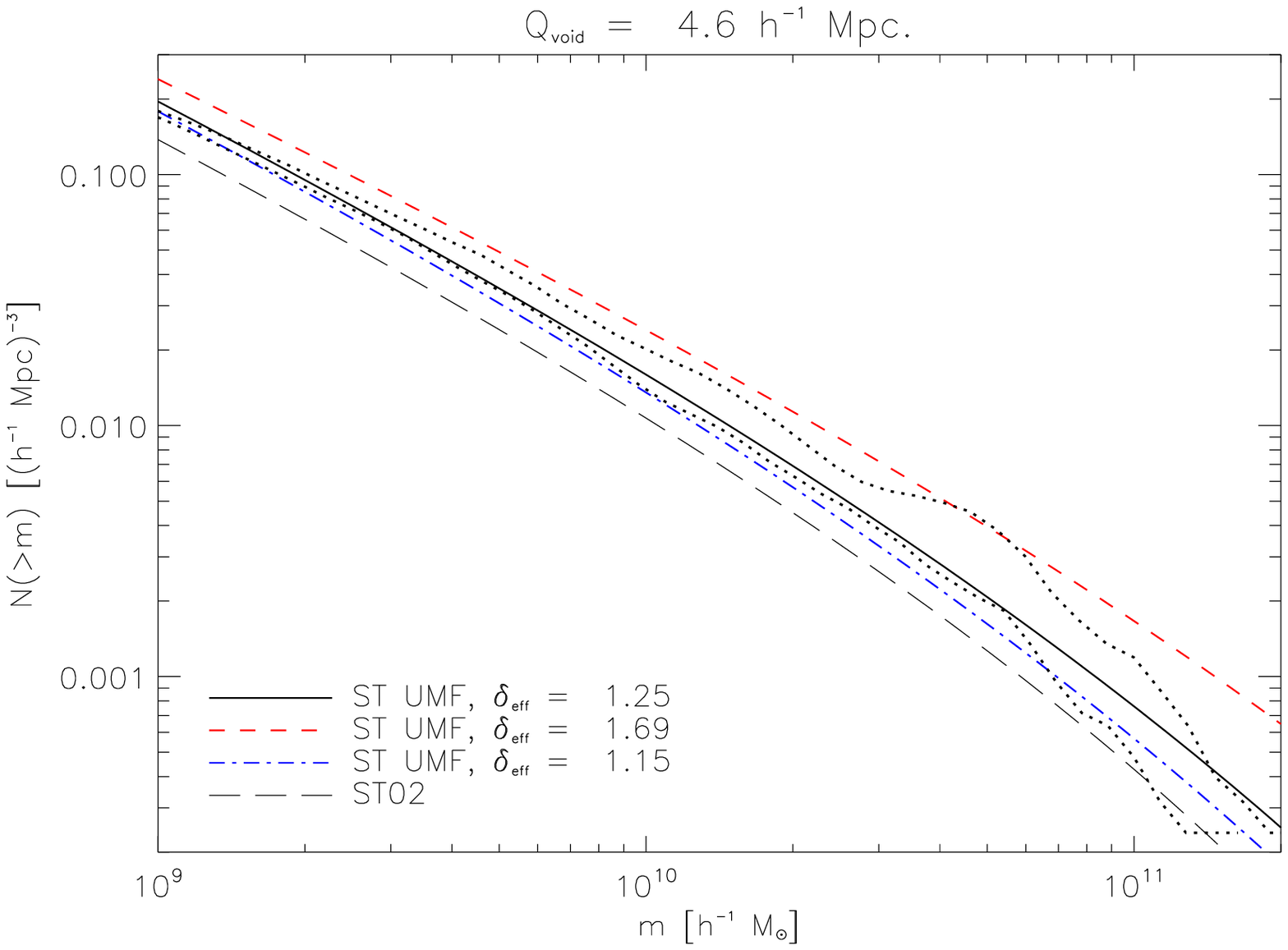}%
\includegraphics[width=\columnwidth]{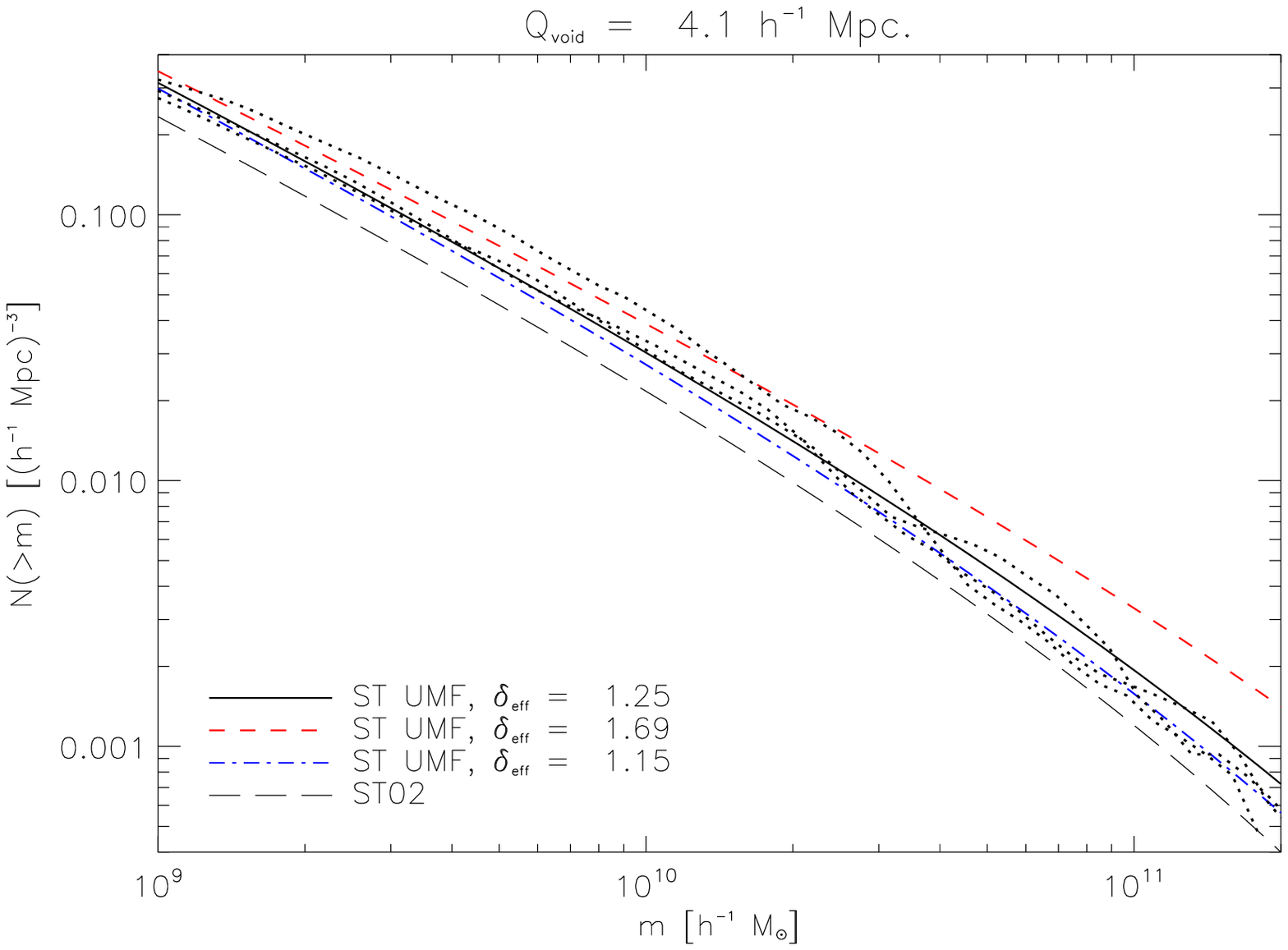}
\caption{ A comparison of the CMF of haloes in voids with the numerical results
  of \citet{Gottlober}. Left panel shows, in dotted lines, two simulated voids
  with $R_{\rm void}=10$~$\runit$ (i.e. $Q_{\rm void} = R_{\rm void}
  (1+\delta)^{1/3} = 4.6~\runit$) and $\deltal = -5.092$. Right panel shows
  three simulated voids with $R_{\rm void}=8$~$\runit$ (i.e. $Q_{\rm void} =
  4.1~\runit$) and $\deltal = -3.995$.  In both panels, the other lines show our
  prescription for the CMF when implemented from the ST UMF.  We consider three
  cases for $\deltaeff$, namely $1.25$ (solid), $1.686$ (dashed), and $1.15$
  (dot-dashed). For comparison, the ST02 mass function is also plotted. It is
  seen that using the standard value of $1.686$ does not reproduce the numerical
  values. However, using the $\deltaeff$ value which fulfils the normalisation
  condition, we correctly reproduce the amplitude and shape of the numerical
  results. }
\label{fig:gott_st}
\end{figure*}

\begin{figure*}
\centering
\includegraphics[width=\columnwidth]{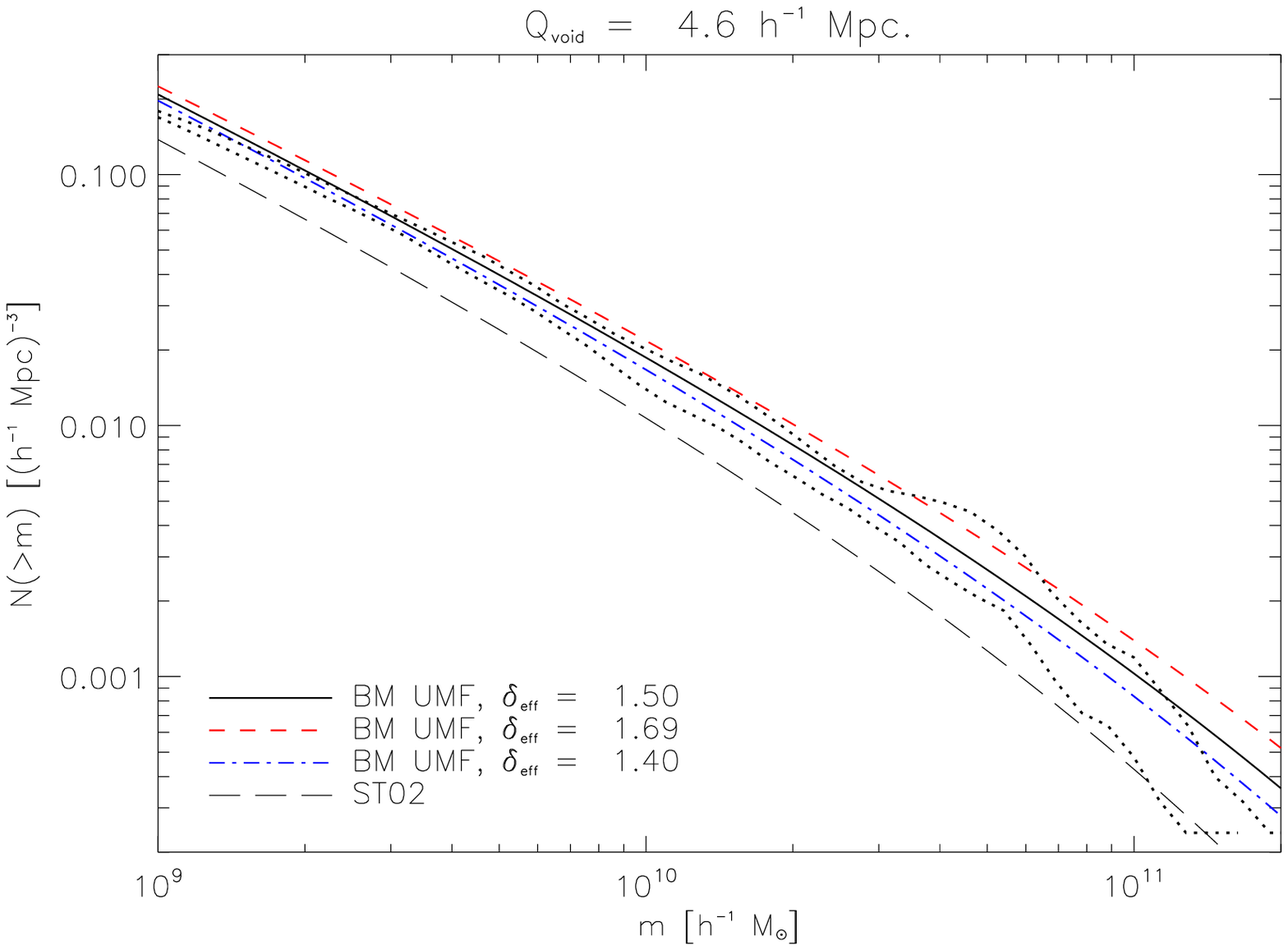}%
\includegraphics[width=\columnwidth]{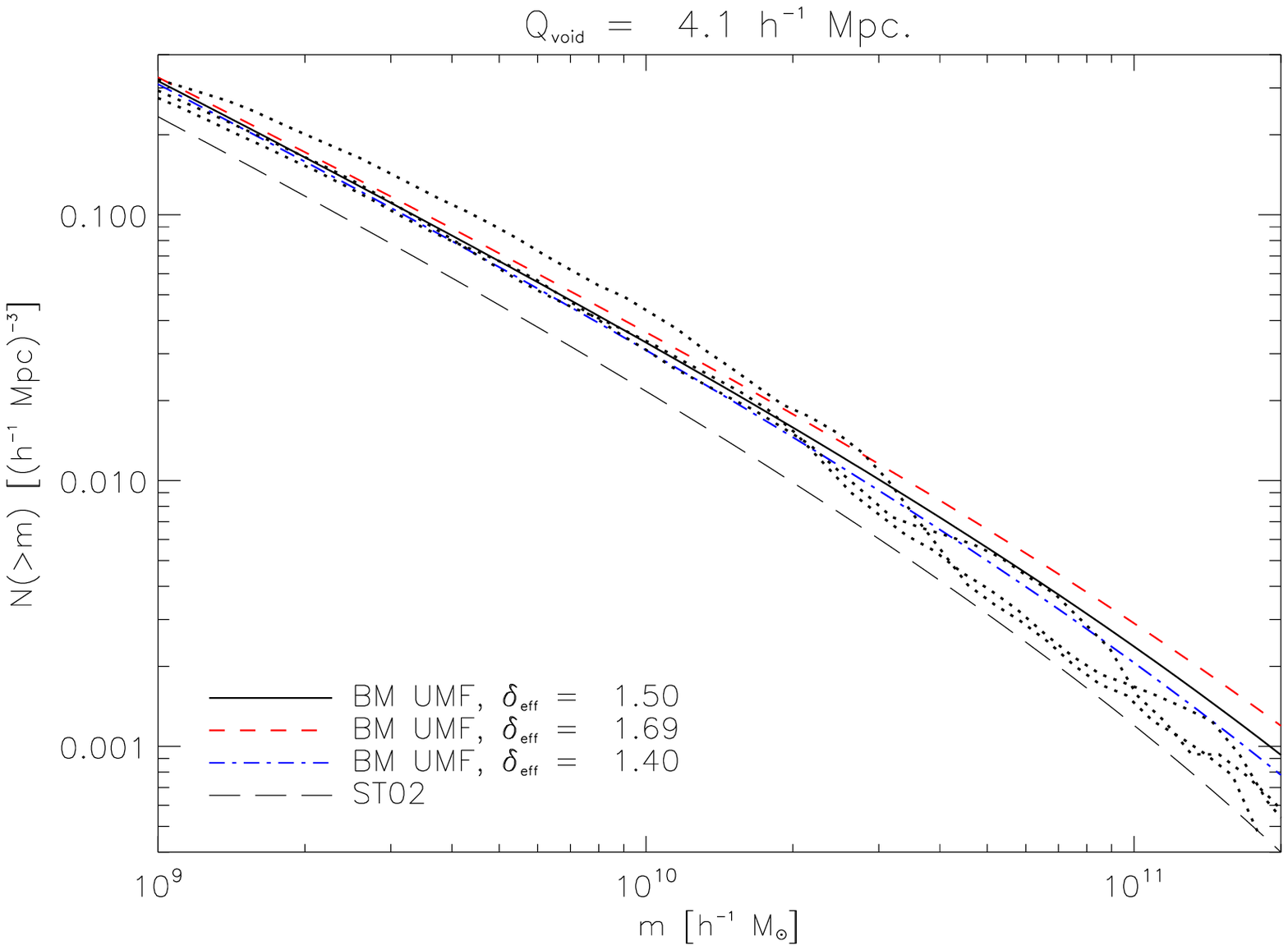}
\caption{ Same as figure~\ref{fig:gott_st}, but using the BM UMF as a reference.
The results are similar to the previous case. Note that for this mass function,
the value of $\deltaeff$ that reproduces the numerical results is closer to
$1.686$. }
\label{fig:gott_bm}
\end{figure*}

\begin{figure*}
\centering
\includegraphics[width=\columnwidth]{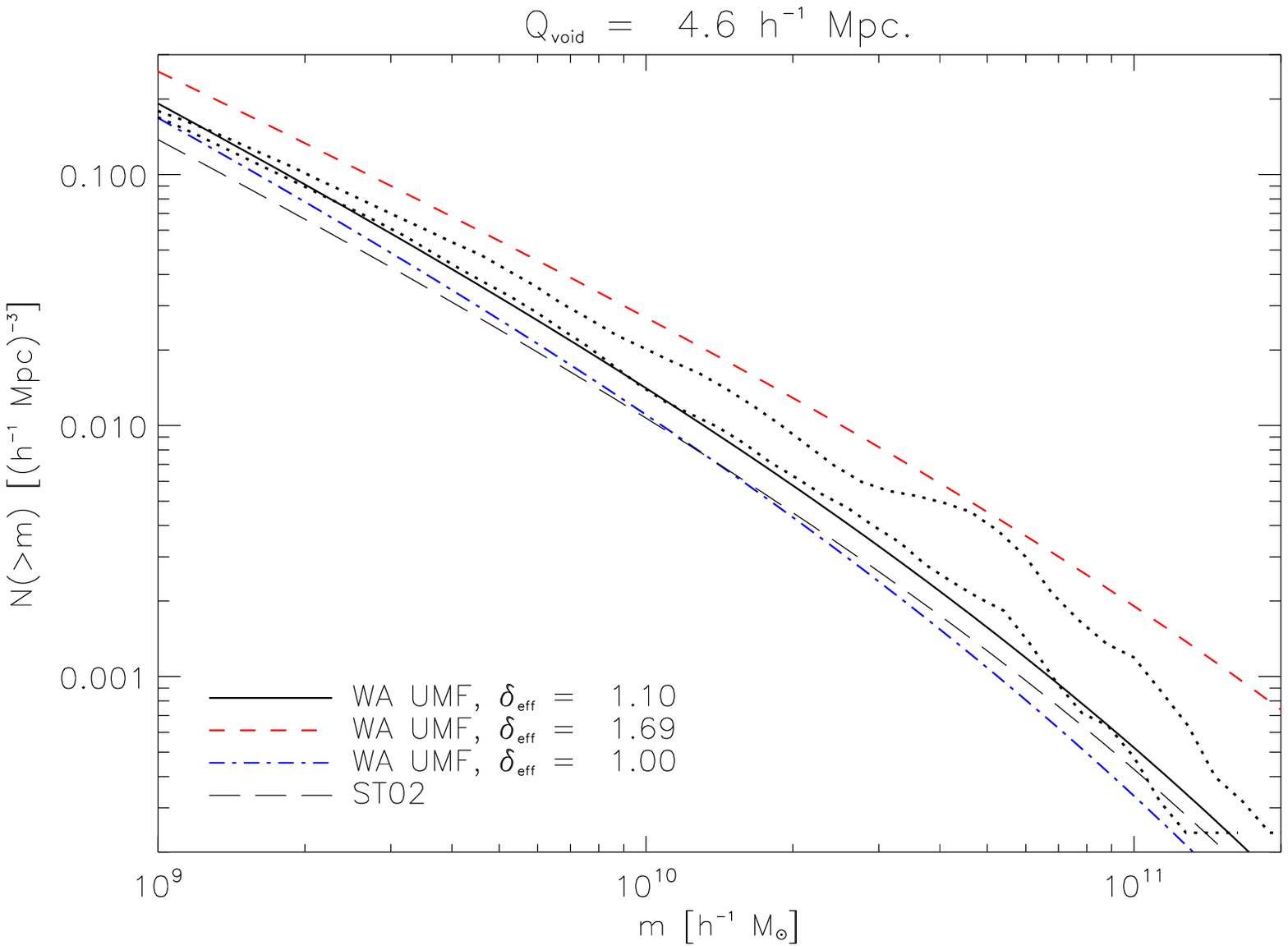}%
\includegraphics[width=\columnwidth]{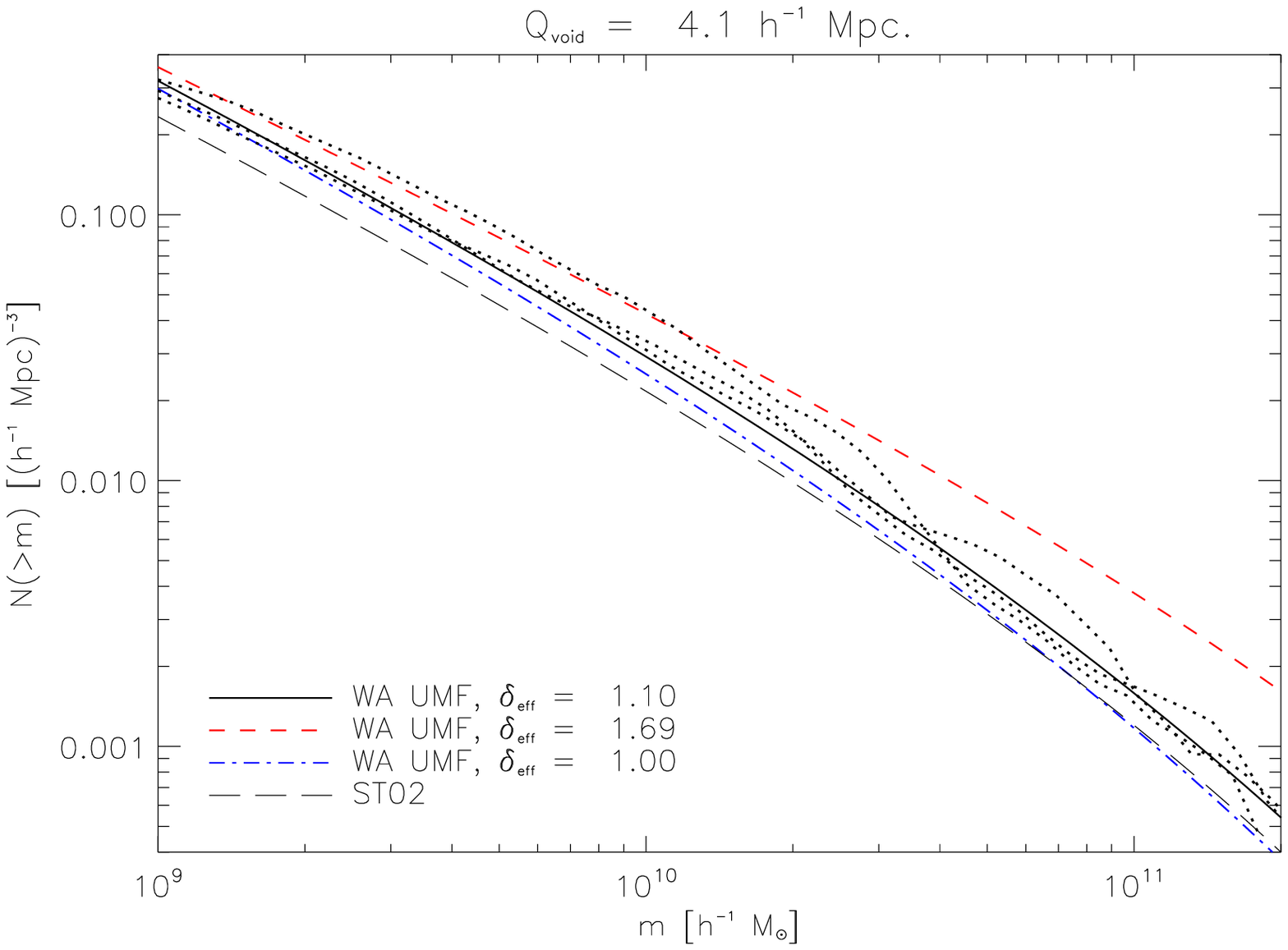}
\caption{ Same as figures~\ref{fig:gott_st} and \ref{fig:gott_bm}, but using the
WA UMF as a reference. Note that for this mass function, the value of
$\deltaeff=1.69$ clearly over-estimates the number of objects at all masses,
specially in the left panel.  }
\label{fig:gott_wa}
\end{figure*}

Figures~\ref{fig:gott_st}, \ref{fig:gott_bm} and \ref{fig:gott_wa} compare the
numerical results with our prediction for the CMF, using three different choices
for the UMF, namely ST, BM and WA. In these figures, it is represented the
accumulated Eulerian mass function averaged within the sphere, which in our
formalism is computed as
\begin{equation}
N_{\rm cE}(m | Q, \deltal) = (1 + \delta_{\rm ns}(m,Q,\deltal)) (1 + \delta(\deltal) )
N(m)
\end{equation}
where we use the $\delta_{\rm ns}$ function computed as in equation~\ref{eq:deltans}.
%

% Comment results.
All three cases are in excellent agreement with the simulations.  Note
that considering the case of $\deltaeff=\deltac \approx 1.69$
overestimates the mass function, specially in the high-mass tail. Note
also that choosing a value for $\deltaeff$ which is 0.1 below the
optimum value, already shows differences, specially in the first set
of simulations ($R_{\rm void} = 10~\runit$).
In addition, the ST02 prescription systematically underestimates these
numerical results by approximately 40\%, specially at low masses where
the numerical uncertainty is much smaller.
The EPS prescription significantly over-estimates these results (by
practically a factor of two), so it is not shown in these figures.
Finally, we stress that there are no free parameters in this
approach. The value for $\deltaeff$ is uniquely determined by the
normalisation condition.

%---------------
%---------------

%---------------
%---------------

%---------------

\section{Discussion and conclusions}

We have developed a procedure to obtain from any analytical expression
for the UMF (either theoretically motivated or numerical fit) a CMF
which is in very good agreement with simulations both for underdense
and overdense conditioning regions.

To do this, we first considered the standard re-scaling (see
eq.~\ref{eq:standard}) in the theoretically motivated expressions for
the UMF (those where $\deltac$ appears explicitly), and pointed out
that to improve the results the re-scaling must be implemented
locally.
We showed that for underdense regions, this leads to a non-negligible
change for masses of the order of $\mstar/100$, while for overdense
regions, the difference is significant only at higher masses.

Next, we showed how to rescale (locally) any UMF even when $\deltac$
does not appear explicitly
(equations~\ref{eq:recipe}-\ref{eq:recipe.v2}).

We then pointed out that those re-scalings are not fully consistent,
because the UMF is not exactly recovered after integrating over all
possible values of the condition.  That is, the ``normalisation
condition'' (eq.~\ref{eq:norma}) is not satisfied. The discrepancy
between both members of this equation may be above 20~\% for the
higher masses, and this is much larger than the precision at which we
know the UMF from numerical simulations ($\sim 5$~\%).

To solve this problem, we modified the rescaling procedure replacing
$\deltac$ by a quantity $\deltaeff$ to be obtained for every UMF by
solving the normalisation condition, which now becomes an equation for
$\deltaeff$. Using these values of $\deltaeff$, all UMFs render very
similar CMFs for both overdense and underdense regions, while for the
former case there is a larger scatter in the mass functions.

Based on these results, we also present (see
appendix~\ref{app4}), an accurate fit for the (accumulated)
conditional mass function, which depends on redshift $z$, and the
cosmology ($\sigma_8$ and $\Gamma$), which can be used in a variety of
problems in large scale structure studies. 

Finally, we point out that based on the derived values for
$\deltaeff$, there is a preference for the BM UMF against the rest of
the UMFs.
Different fits lead to somewhat different values of $\deltaeff$, but
all of them are incompatible with theoretically motivated values which
must satisfy $\deltaeff \ga 1.6$. However, if we apply locally the
re-scaling procedure that follows from the approach used by
\citet{BM}, which involves re-scaling certain local spectral constant
in addition to the mean and variance of the density field, the
resulting CMF is also in very good agreement with the simulations but
now the value of $\deltaeff$ is theoretically ``acceptable'' (note
that the range of 5\% is the only one which is compatible with
$\deltaeff \approx 1.6$). This suggests that this last rescaling is
the ``correct one'', although most of the results in this work may be
equally well obtained with any other prescription. 
%

%%%%%%%%%%%%%%%%%%%%%%
%%
\section*{Acknowledgments} 
We thank Stefan Gottl\"ober for kindly providing us in electronic form
the statistics of voids from the simulations described in Gottl\"ober
et al. (2003).

%%%%%%%%%%%%%%%%%%%%%%
%%

%%%%%%%%%%%%%%%%%%%%%%%%%%%%%%
%
\appendix

\section{Analytic expressions for the mass fraction of dark matter haloes}
\label{app1}

For some of the computations in this paper, we found that it is useful to have
an analytical expression for the mass fraction $F(m)$. We present here these
expressions for the case of \cite{PS}, \cite{ST} and \cite{WA} mass functions. 

For the case of the PS mass function, we have the well known result of
\begin{equation}
F_{\rm PS}(m) = {\rm erfc} \Big( \frac{\delta_c}{ \sqrt{2} \sigma(m) } \Big)
\end{equation}
where ${\rm erfc}$ is the complementary error function, defined as ${\rm
erfc}(x) = 2/\sqrt{\pi} \int_{x}^{+\infty} e^{-t^2} dt$.

\noindent
For the case of the ST mass function given in equation~\ref{eq:f_st}, we can
integrate eq.~\ref{eq:F_and_f} and obtain
\begin{equation}
F_{\rm ST}(m) = A \Big[ 
{\rm erfc} \Big( \frac{\sqrt{a} \delta_c}{ \sqrt{2} \sigma(m) } \Big) + 
\frac{1}{\sqrt{\pi} 2^p } \Gamma( \frac{1}{2} - p, \frac{a}{2} ( \frac{\delta_c}{\sigma} )^2 )
\Big]
\end{equation}
where $\Gamma(\alpha,x)$ represents the incomplete Gamma function,
defined as $\Gamma(\alpha,x) = \int_{x}^{+\infty} e^{-t} t^{\alpha-1} dt$. 

\noindent
Finally, for the case of the WA mass function given by equation~\ref{eq:f_wa}, we
obtain
\begin{equation}
F_{\rm WA}(m) = \frac{1}{2} A \Big[
c^{-a/2} \Gamma( \frac{a}{2} , \frac{c}{\sigma^2} ) + b {\rm E}_1( \frac{c}{\sigma^2} )
\Big]
\end{equation}
where ${\rm E}_1(x)$ is the exponential integral function, defined 
as ${\rm E}_1(x) = \int_{x}^{+\infty} e^{-t} dt/t$.

\section{Some useful analytic fits to the $D$ function}
\label{app2}

We present here some useful fits to the $D(q,Q,Q_2)$ function, which was defined
in equation~\ref{eq:Dq}. All the numbers in this appendix correspond to the
transfer function and cosmological parameters presented at the end of Sec.~1
(see equation~1).

In most of the cases considered in this paper, we are dealing with scales which
are much smaller than the condition (i.e. $Q_2 \ll Q$). In that limit, the
$D(q,Q,Q_2)$ function is practically independent on $Q_2$, and the following
fitting function reproduces well the overall shape
\begin{equation}
D(q,Q,Q_2) = A(Q) e^{ -B(Q) (\frac{q}{Q})^2}, \qquad Q_2 \ll Q
\end{equation}
Note that in the limit $q\rightarrow 0$, we would expect this exponential
dependence to be exact. 

A numerical fit in the scale range $4$~$\runit$~$< Q <$~$13$~$\runit$ gives the
following values:
\begin{equation}
A(Q) = 1.264 + 0.167 Q_8 -0.0415 Q_8^2, 
\end{equation}
\begin{equation}
\ln B(Q) =  -0.5632 + 0.1787 (\ln Q_8) -0.0222 (\ln Q_8)^2 
\end{equation}
where $Q_8 = Q/8$~$\runit$. The typical error of this fit with respect to the
exact computation is of the order of 5\%.

%%%%%%%%%%%
% At the center
It is also interesting to provide a fit for the $D$ function evaluated at the
center of the condition, $q=0$. In that case, we obtain
\begin{equation}
D(q=0,Q) = 1.262 + 0.1522 Q_8 -0.0392 Q_8^2.
\end{equation}

Finally, it is also useful to fit the function $c(m)=D(Q,Q,Q)$, where $m=m(Q)$. In this case,
considering the mass range $10^8$~$\munit$ $< m < 5\times 10^{15}$~$\munit$, we
obtain
\begin{equation}
c(m) = \sum_{i=0}^{i=4} c_i (\ln m)^i,
\end{equation}
with $(c_0,c_1,c_2,c_3,c_4) = 
(1.06004, -0.16521, 0.01024, -2.72\times10^{-4}, 2.96\times10^{-6})$.

\section{Analytic fits to the $\delta_{\rm ns}$ function}
\label{app3}

In this section we provide analytic fitting formulae to the
$\delta_{\rm ns}$ function, described in equation~\ref{eq:deltans}. To
a good approximation, this function can be parameterised in terms of
two functions, $A(m,Q)$ and $b(m,Q)$, which are defined in
equation~\ref{eq:Ab}.

All the following fits have been obtained with the ST UMF and using
$\deltaeff=1.25$.  We have considered values of $Q$ within the range between 5
and 13~$\runit$.
The mass interval for all fits is taken to be $10^9 \munit < m < m(Q)/30$.
However, we find that the functional form proposed in equation~\ref{eq:Ab} only
provides a reasonable fit (i.e. with errors of few percent) to the data for
masses above $m/\mstar \ga 3\times 10^{-3}$ (where $\mstar$ is defined as
$\sigma(\mstar)=1$).  For mass values outside that range, it is necessary to do
the numerical integration in Eq.~\ref{eq:deltans}.
Finally, all these fits have been obtained for values of the linear density
within the range $ -4.5 < \deltal < -1$. These are the typical values we are
interested in. We note that beyond this range, and in particular, for values of
$\deltal$ close to zero, these fits may give inaccurate results.

The proposed fitting formula for $A(m)$ is simply a quadratic fit in $\ln m$,
i.e.
\begin{equation}
A(m,Q) = \sum_{i=0}^{i=2} a_i(Q) (\ln m')^i 
\label{eqapp:A}
\end{equation}
where $m'$ is the dimensionless mass given by
\begin{equation}
m' = \frac{m}{3.51\times 10^{11}~\munit}
\end{equation}
and the coefficients are linear functions of $Q$,
\begin{equation}
a_i(Q) = a_i^0 + a_i^1 \Big( \frac{Q}{8~\runit} \Big)
\end{equation}
The results we obtain are 
$(a_0^0,a_0^1) = (1.577,-0.298)$,  
$(a_1^0,a_1^1) = (-0.0557, -0.0447)$, and 
$(a_2^0,a_2^1) = (-0.00565, -0.0018)$. 

The best fit for $b(m)$ is obtained with the formula proposed in \cite{Patiri}, 
\begin{equation}
b(m,Q) = b_1(Q) + b_2(Q)~(m')^{b_3(Q)}
\label{eqapp:b}
\end{equation}
but the coefficients are now linear functions of $Q$, 
\begin{equation}
b_i(Q) = b_i^0 + b_i^1 \Big( \frac{Q}{8~\runit} \Big)
\end{equation}
Using these expressions to fit the numerical results, we find 
$(b_1^0,b_1^1) = (-0.0025, 0.00146)$, 
$(b_2^0,b_2^1) = ( 0.121, -0.0156)$, and 
$(b_3^0,b_3^1) = (0.335, 0.019)$.

\section{Analytical fit to the conditional mass function for any cosmology}
\label{app4}

Following the approach presented in this paper, we give here
a highly accurate analytical fit to the CMF, as a function of the
redshift and the cosmology.
The (accumulated) conditional mass function $N_{\rm c}(m)$, as a
function of the linear density contrast $\deltal$, within a spherical
conditioning region with \emph{Lagragian} radius $Q$ is given, for any
value of the redshift $z$, and the cosmology ($\sigma_8$ and
$\Omega_{\rm m}$), by

\begin{equation}
N_{\rm cE}(m,z | Q, \deltal ; \sigma_8, \Gamma) = \mathbf{A}
e^{-\mathbf{b} \deltal^2} ( 1 + \delta ) N(m,z),
\end{equation}
where $N(m,z)$ is the (unconditional) accumulated mass function,
$\delta$ is the actual fractional mass density fluctuation, and the
coeffitients $\mathbf{A}$ and $\mathbf{b}$ are functions of $m$, $Q$,
$z$, $\sigma_8$ and $\Gamma$ given by
\begin{equation}
\mathbf{A}(m,Q; z, \sigma_8, \Gamma) = \Bigg(\frac{\sigma_8
b(z)}{0.9}\Bigg)^{0.71 + 0.08m'} A(m,Q)
\label{app4:eq1}
\end{equation}
\begin{equation}
\mathbf{b}(m,Q; z, \sigma_8, \Gamma) = \Bigg(\frac{\sigma_8
b(z)}{0.9}\Bigg)^{-2.65} b(m,Q)
\label{app4:eq2}
\end{equation}
and where $b(z)$ is the growth factor of linear density perturbations normalised
to one at $z=0$, and $A(m,Q)$ and $b(m,Q)$ are given by equations~\ref{eqapp:A} and
\ref{eqapp:b}, respectively, but using this new mass definition 
\begin{equation}
m' = \frac{m}{3.51\times 10^{11}~\munit} \Bigg(\frac{0.21}{\Gamma}\Bigg)
\end{equation}
It is found that the $\Gamma$ dependence of both $\mathbf{A}$ and
$\mathbf{b}$ is negligible, but the dependence on $\sigma_8$ (and
therefore on redshift $z$) is rather strong, particularly in the case
of $\mathbf{b}$. 
These two scalings presented in equations~\ref{app4:eq1} and
\ref{app4:eq2} have been obtained as a fit to the numerical results
for values of the linear density in the range $-4.5 <
\deltal(0.9/\sigma_8) < -1$, as in the previous appendix.  We have
checked that changing the range of integration may produce changes on
the exponent of these scalings up to the order of 10 per cent.
The exponent in both cases has been fitted in the mass range $m >
3\times 10^{-4}~\mstar$.

Finally, it must be noted that the range of values of $\sigma_8$ and
$\Gamma$ used to obtain this fit is roughly a factor two around the
reference model (i.e. $\sigma_8= 0.9$ and $\Gamma =0.21$). Thus,it can
not be safely extrapolated beyond $z\approx 2$. Furthermore, for large
redshift values ($z > 5$), the considerations contained in \citet{BM2}
must be taken into account.

\label{lastpage}
\end{document}